\documentclass[acmsmall,screen,nonacm]{acmart}
\setcopyright{none}
\usepackage{booktabs}
\usepackage{tabularx}
\usepackage{listings}
\usepackage{tikz}
\usetikzlibrary{positioning,arrows.meta}
\newcolumntype{Y}{>{\raggedright\arraybackslash}X}
\lstdefinelanguage{vx}{
  morekeywords={fn,let,mut,for,in,if,else,loop,while,break,continue,return,spawn,tasks,on,
    where,transfer,across,match,struct,unsafe,extern,pub,trait,type,impl,comptime,import,
    assert,assume,enum,as,requires,ensures,invariant,Memory,Topology,mlir,void,safe,const},
  sensitive=true, morecomment=[l]{//}, morestring=[b]"}
\newcommand{\kw}[1]{\texttt{#1}}
\newcommand{\st}[1]{\textsf{\small[#1]}}

\begin{document}

\title{Vx: A Programming Language for Scalable, Maintainable and Secure Systems}

\author{Aditya Kumar}
\email{hiraditya@msn.com}
\affiliation{%
  \position{Principal Engineer}
  \institution{Independent}
  \country{USA}}

\begin{abstract}
A program for a modern machine runs on several kinds of chips. Its memory spaces, links and capacities are invisible to the compilers of today. So a host that reads
device memory, a working set that does not fit, or a data race shows up only when the program runs. Vx is a
systems programming language that puts these facts in the program text. A machine is a source
file of declarations. A value's type says which memory it lives in. Apart from one stated exception, a function's signature is the whole contract that its callers
see. A flag lists the code and the facts in a program that the compiler trusts without a proof. This report is a tour of the language at one commit. It describes six design
principles, the core language, and the features that serve three qualities. Vx is scalable because one program text runs across machines, cores and GPU threads.
With an option, the compiler also checks its function bodies in parallel. It is maintainable because a signature is the contract of a function and most diagnostics
have codes. The output of the parallel compiler does not depend on the number of threads. It is secure
because safe code has ownership, borrow, race and placement checks. In safe code, the standard library
asks for a right to open a file or a socket, with a few listed exceptions. For every feature, the report gives its status and the evidence for
it. It also lists where the implementation does not yet meet the design.
\end{abstract}

\keywords{systems programming language, heterogeneous computing, ownership, data-race freedom,
compile-time evaluation, contracts, parallel compilation, machine models}

\maketitle

\section{Introduction}\label{sec:intro}

A large program of today does not run on one processor. A model server keeps its weights in
GPU memory, stages requests through host memory, and sends work to other machines. A
simulation splits its loops over CPU cores and GPU threads. Each part of such a program has
its own memory spaces, its own links between them, and its own limits.

The languages that these programs are written in do not see those facts. In C++ and CUDA, a
pointer into GPU memory has the same type as a pointer into host memory. A capacity is a
number in a configuration file or in the head of a programmer. So the errors appear late.

A host read of a device pointer crashes. A working set that does not
fit fails after someone rents the hardware. A data race corrupts a result on one run in a thousand. Rust removes
data races from safe code~\cite{jung2018rustbelt}, but it also has no type for ``this value lives
in GPU memory''.

Vx~\cite{vx2026} is a systems programming language that puts these facts in the program text.
Its central idea is that \emph{the machine is part of the program}. A machine is a source file
that declares memory spaces, capacities, links and execution levels. A value's type says where
it lives, so \kw{Tensor<f32, [4, 4], Memory::GPU\_HBM>} and \kw{Tensor<f32, [4, 4]>} are
different types. Code that a device cannot address does not compile against it, and the error
names the transfer that fixes it.

\paragraph{Three qualities.} The title names three qualities. This report uses each word in a
narrow sense, and it gives the evidence for each claim.
\begin{itemize}
\item \textbf{Scalable.} One program text runs across machines, CPU cores and GPU threads. The
  compiler checks the placement of each value against each declared machine. The compiler also
  scales with the size of a program. It can check function bodies in parallel, because no body
  adds a fact that another body needs.
\item \textbf{Maintainable.} A reader can understand a function from its signature and its
  body. There is no overloading, no implicit conversion and no hidden data movement. Most diagnostics have a code, and each code has a page. The compiler output does not depend on the number of
  threads, and a tool reports how a library interface changed between two versions.
\item \textbf{Secure.} Safe code has ownership and borrow checks and a check that tasks do not
  race. Other checks make sure that a device reads only memory that it can address. Integer overflow has a
  defined result, and contracts are checked with an encoding that knows about it. In safe code,
  \kw{File::open} and the socket calls of the standard library need a right, which a function gets
  from its caller. Section~\ref{sec:authority} lists the exceptions. The compiler trusts some code
  and some facts without a proof, and a flag lists those of the program.
\end{itemize}

\paragraph{Status of a young language.} Vx is an early research language. Its first commit is
from May 2026, and this report describes the main branch at commit \kw{719726333} of
23 November 2026 (version 0.0.5). A tour of a young language can easily claim more than the
compiler does. So each feature in this report carries one status word:
\begin{itemize}
\item \st{tested}: the code exists, and continuous integration (CI) tests it;
\item \st{tested by hand}: it ran on real hardware, such as a GPU, but no CI job runs it;
\item \st{partial}: some of the described behavior works;
\item \st{designed}: a document describes it, but the compiler does not do it yet.
\end{itemize}

Section~\ref{sec:secure-gaps} lists the places where safe code is not yet safe, and
Section~\ref{sec:status} collects the status of every feature in one table.

\paragraph{Contributions.} This report makes four contributions.
\begin{enumerate}
\item It describes the whole language at one commit: the core language and the features
  for scale, maintenance and security (Sections~\ref{sec:core}--\ref{sec:secure}). It also describes the
  implementation (Section~\ref{sec:impl}).
\item It states six design principles and the rules of the language that make each one hold. For
  each principle, it names one place where the implementation does not yet follow it
  (Section~\ref{sec:principles}).
\item It gives the status and the evidence for each feature: the tests, the hardware, and the
  formal models (Section~\ref{sec:status}).
\item It gives an account of what is not yet safe or not yet built, with the issue that tracks
  each gap (Sections~\ref{sec:secure-gaps} and~\ref{sec:limits}).
\end{enumerate}

\paragraph{Companion papers.} Several parts of Vx have papers of their own, and this report
points to them for detail. The fleet-level report describes machines, placement and
\kw{across} loops for multi-node work~\cite{kumar2026fleet}. Must-Fold and the comptime report
describe compile-time evaluation and its soundness~\cite{kumar2026mustfold,kumar2026comptime}.
Structured Tasks and Footprints describes the concurrency model and its Lean
proofs~\cite{kumar2026tasks}. Erasable Parallelism describes GPU loop
nests~\cite{kumar2026erasable}.

Agree Without Locking describes the parallel compiler
frontend~\cite{kumar2026frontend}. A study of machine models tests the machine files against
rented hardware~\cite{kumar2026machinemodel}. A study of LLM serving tests capacity verdicts on
real GPUs~\cite{kumar2026admission}.

\paragraph{A first program.} Listing~\ref{lst:first} shows the shape of a Vx program. It makes
two matrices in host memory, moves them to the memory of an accelerator, multiplies them there,
and moves the result back.

\begin{lstlisting}[float=t,caption={A first program. Each move between memory spaces is an
explicit \kw{transfer}, and \kw{spawn on} names the device that runs the block.},label=lst:first]
fn main() -> i32 {
  let a = Tensor<f32, [4, 4]>::fill(2.0);
  let b = Tensor<f32, [4, 4]>::fill(3.0);
  let mut c = Tensor<f32, [4, 4], Memory::NPU_HBM>::new();
  let a_npu = transfer(a, Memory::NPU_HBM);
  let b_npu = transfer(b, Memory::NPU_HBM);
  spawn on(Topology::NPU[0]) {
    c = a_npu @ b_npu;
  }
  let c_host = transfer(c, Memory::CPU_DRAM);
  print(c_host[0][0]);   // 24
  return 0;
}
\end{lstlisting}

If the program uses \kw{a} inside the block and drops the line that makes \kw{a\_npu}, the
compiler refuses it:
\begin{lstlisting}[language={},basicstyle=\ttfamily\scriptsize]
Error[E6003] at first.vx:7:9: 'a' lives in CPU_DRAM but NPU[0] sees only [NPU_HBM];
insert an explicit transfer to NPU_HBM (cost 50 on the declared path)
\end{lstlisting}
The compiler knows that the device cannot address host memory, because the built-in NPU topology
says that the device sees only \kw{NPU\_HBM}. It also finds the cheapest declared route and its cost. On the Apple M4 where we ran this
program, the runtime found no prebuilt model for the neural engine. It printed that the block runs on the CPU, and the result was 24. The shipped M4 machine file
refuses this program, because its neural engine declares no \kw{f32} type (E6026). Section~\ref{sec:backends} says what runs on each
target.

\section{Design Principles}\label{sec:principles}

Six principles shape Vx. For each one, this section gives the rules of the language that make it
hold. It also names one place where the implementation at \kw{719726333} does not yet follow it. The
counterexamples are not a side note. A principle is useful only if a reader can test the code
against it.

\subsection{P1: The machine is part of the program}

A Vx program states where each value lives and which device runs each block. The project's
manifesto puts it in one line: ``Heterogeneity belongs in the type system, not in the runtime.''
Three rules make this hold.
\begin{itemize}
\item A machine is a source file of declarations (Section~\ref{sec:machines}). The compiler
  reads it with \kw{--machine}, and the program does not name the file.
\item A value's type carries its memory space. A tensor type has an element type, a shape and a
  space; a raw pointer can name its space too (\kw{*mut f32 in Memory::GPU\_HBM}).
\item Data moves only through an explicit \kw{transfer}. This holds even on a unified-memory chip. There, the copy costs nothing in the model, and the
  compiler skips it when each use only reads. The move is still written, so a reader sees
  where the program moves its data.
\end{itemize}

\emph{Where the code does not yet follow it:} the compiler reads one machine file for each compilation, and a machine file cannot import
another. So the node files copy the declarations of one GPU. The eight-GPU file puts its seven
peers in one 560~GiB space, so it admits placements that no single peer can hold.

\subsection{P2: A signature is the whole contract}

A caller of a Vx function needs only the function's signature. This principle serves both scale
and maintenance. It lets the compiler check function bodies in parallel, and it lets a
reader understand a call without the body of the callee. The project states it as eight rules,
each a rule that C++ or Rust breaks (Table~\ref{tab:freeze}). Before the compiler checks any
body, it collects every global fact (every signature, layout and impl) and freezes it. A body
only reads the frozen table.

\begin{table}[t]
\caption{The eight rules that let the compiler freeze every global fact before it checks any
function body, from \kw{docs/language\_rules\_for\_a\_parallel\_compiler.md}, with a feature of
C++ or Rust that breaks each rule.}\label{tab:freeze}
\small
\begin{tabularx}{\linewidth}{@{}p{0.36\linewidth}Y@{}}
\toprule
Vx rule & A feature that breaks it \\
\midrule
A signature is the whole contract: every parameter and result type is written. &
  C++ \kw{auto f()}; Rust \kw{-> impl Trait} and \kw{async fn}, where a property of the result
  comes from the body. \\
Every layout is known at the freeze. & Rust \kw{[u8; n()]}: a size comes from running a
  \kw{const fn}. \\
A generic closes at its definition. & C++ specializations after use and argument-dependent
  lookup; Rust specialization. \\
One name, one function: no overloading. & C++ open overload sets. \\
No implicit conversions. & C++ converting constructors and \kw{operator T()}. \\
Copy and drop belong to the type. & C++ special members that are made on first use. \\
Macros expand in their own phase. & The C preprocessor; Rust macro expansion interleaved with
  name resolution. \\
A body declares nothing global. & Rust \kw{impl Trait for S} inside a function body. \\
\bottomrule
\end{tabularx}
\end{table}

One global fact can still appear inside a body: a new instantiation of a generic. The compiler
gives it an identity that does not depend on which thread finds it first
(Section~\ref{sec:frontend}).

\emph{Where the code does not yet follow it:} Vx signatures have no
lifetimes. So the compiler reads one fact about a result from the body of the callee. That fact is
where a returned reference may point (Section~\ref{sec:ownership}). The compiler reads it from the
syntax of every body before it checks any body, so the check stays parallel. But a change to a
body can change which callers compile, and the documentation says so.

\subsection{P3: Refuse rather than guess}

When the compiler cannot decide a fact, it refuses the program. It does not choose a likely
answer.
\begin{itemize}
\item A \kw{comptime} block that the compiler cannot finish is an error (E3033). There is no
  silent fallback to run time (Section~\ref{sec:comptime}).
\item Contracts need the Z3 solver. When Z3 is missing, each contract fails, with a warning that
  says why (Section~\ref{sec:contracts}).
\item The reader of a compiled library interface fails closed. It skips a signature that it
  cannot encode, and it checks the bounds of every read.
\item A device that cannot address a space does not get a silent copy. The program gets E6003,
  and the message names the transfer.
\end{itemize}

\emph{Where the code does not yet follow it:} run-time bounds checks are off by default. So the
program does not check an index at run time when the compiler cannot check it at compile time
(Section~\ref{sec:secure-gaps}). Also, most regions that a program places on the Apple neural engine run on the CPU with no
message (\#3266). One route picks a model by the shapes of the buffers alone, and it gives wrong results (\#3793,
Section~\ref{sec:backends}).

\subsection{P4: Every number has a source}

A capacity verdict is only as good as the capacity that it uses. So each figure in a machine file
carries a comment that names its source. The source is a vendor document (\kw{spec:}) or a
measurement (\kw{measured:}). A figure with no source should be marked \kw{UNVERIFIED}. The 16 machine files that ship
with the compiler have 126 \kw{spec:} comments, 6 \kw{measured:} comments and 42 \kw{UNVERIFIED}
marks. Units are exact: \kw{GB} is $10^9$ bytes and \kw{GiB} is $2^{30}$, so a figure copied from a
vendor sheet means what the sheet meant.

A machine-model study rented the hardware, measured the
figures, and recorded where the declarations were wrong~\cite{kumar2026machinemodel}. The
project's benchmark policy applies the same rule to performance claims. It asks for absolute
numbers, the best baseline and at least three runs, and it puts the losses in the same table as
the wins.

\emph{Where the code does not yet follow it:} the provenance comments are a convention. No tool reads them, and CI does not check that every
figure has one. In the H100 file, some on-chip figures have neither a source nor a mark. The Intel
Arc file declares 1,024 lanes in a block. Its comment says that nobody measured this figure, but the
figure has no \kw{UNVERIFIED} mark.

\subsection{P5: One meaning on every target}

A Vx program means the same thing on every target, and the project tests this with a second
version of the same program.
\begin{itemize}
\item An \kw{across} loop runs its passes in parallel on a GPU. Delete each \kw{across}, and the loops run in order with the same result, unless the
  kernel calls a \kw{lanes::} routine (Section~\ref{sec:across}). The compiler also makes a
  CPU copy of each kernel, and a test can compare the two.
\item A \kw{comptime} block gives the value that the same code gives at run time. A test suite
  runs each program as written, and again with \kw{comptime} replaced by a plain block, and the
  outputs must match.
\item The compiler has two code generators, and a differential fuzzer runs each generated program
  through both of them and through an equivalent Rust program.
\item With \kw{-j}, the compiler output is the same bytes at 1, 4 and 8 threads
  (Section~\ref{sec:determinism}).
\end{itemize}

\emph{Where the code does not yet follow it:} the two code generators still disagree on some
programs. The job that compares them on the whole test suite began on 10 November 2026, and each
of its four runs to 23 November failed.

\subsection{P6: Say what is trusted, and where each promise stops}

Every language trusts some code. Vx tries to say which code. An \kw{unsafe} block, an
\kw{unsafe fn}, an \kw{unsafe impl} and an \kw{extern impl} mark code whose claims the compiler
cannot check. A foreign function is an unchecked promise. An \kw{mlir!} block outside \kw{unsafe}
must be one that the compiler can verify. The flag \kw{--audit} lists the trusted code and the
trusted facts of the program, and it counts those of \kw{core} and \kw{std}. The flag \kw{--no-trusted-code} refuses trusted code in a package
(Section~\ref{sec:trusted}).

The book's
page on concurrency states each promise, then says where it stops. It also says how each promise
is checked: by a proof about a model, by a model checker on compiled code, or by tests only. The manifesto states
the rule as a refusal: Vx ``will not claim what it has not checked''.

\emph{Where the code does not yet follow it:} a foreign function that is declared \kw{safe} needs
no \kw{unsafe} at its calls. So a file can declare the C function \kw{system} as \kw{safe} and run a
shell command from safe code. \kw{--audit} lists the declaration, but each call looks like the call
of a safe Vx function (Section~\ref{sec:authority}).

\section{The Core Language}\label{sec:core}

This section is a tour of the parts of Vx that a program for one CPU uses. Rust is the nearest
relative, so the tour often says how Vx differs from Rust. Each subsection ends with the status of
its features.

\subsection{Items, types and numbers}\label{sec:items}

A Vx file is a module. It declares functions, structs, enums, traits, impls, constants and
macros. A function writes the type of every parameter and of its result: \kw{-> void} when there
is no result, and \kw{-> !} when it never returns. A \kw{const} needs a type. There is no
\kw{static}, no \kw{union}, no inline assembly and no attribute syntax.

The integer types are signed and unsigned, from 4 to 64 bits (\kw{i4} \ldots
\kw{i64}, \kw{u4} \ldots \kw{u64}), and \kw{usize}. The floating-point types are \kw{f16},
\kw{bf16}, \kw{f32} and \kw{f64}, plus eight-bit and four-bit formats for machine learning.
Vector types are written \kw{<N x T>}. There are no implicit conversions. Every conversion is an explicit \kw{as}, and a
number literal takes its type from the place where the program uses it. A literal that does not fit
its type is an error (E3045).

Integer arithmetic has a defined result for every input:
\begin{itemize}
\item \kw{+}, \kw{-} and \kw{*} wrap at the width of the type, as in two's complement hardware;
\item \kw{/} and \kw{\%} by zero stop the program, and so does the minimum value divided by $-1$;
\item arithmetic on \kw{usize}, which holds 0 to the maximum of \kw{i64}, stops the program when
  it overflows;
\item \kw{as} wraps, but a cast to \kw{usize} out of range stops the program. When the compiler
  knows the value, it also gives the warning W1014.
\end{itemize}

A shift by the width of the type or more is the one undefined case, and the documentation says so.
The contract checker uses the same wrap-around rules (Section~\ref{sec:contracts}), so an overflow
cannot make a false postcondition look true.

\emph{Status:} \st{tested}. The 128-bit integers are \st{partial}: arithmetic works, but
\kw{print} fails in the code generator.

\subsection{Data: structs, enums, tuples and tensors}\label{sec:data}

A struct has named fields, and a struct literal writes them in the order of the declaration
(E3062). An enum is a tag and a payload, and a variant always writes its type, as in
\kw{Option<i64>::Some(3)}. A tuple \kw{(A, B)} is a library struct \kw{Tuple2<A, B>}, with
versions for two to six elements, and the parser rewrites tuple patterns to it.

Vx has no separate array type. A fixed-size array is a tensor: \kw{Tensor<i64, [4]>} holds four
integers, and \kw{Tensor<f32, [64, 300]>} is a matrix. A dimension can be a number, a
\kw{const} generic, or \kw{?} for a size that is known only at run time. Rows, ranges of rows,
reshapes and fields of tensors are views, and the borrow checker treats each view as a borrow.
There is no slice type \kw{\&[T]}. A recursive data type goes through a \kw{Box}, as in
\kw{Option<Box<Node>>}.

A \kw{match} takes enum variants (with a name or \kw{\_} for the payload), literals, tuples and
\kw{\_}. The compiler checks that a match on an enum or a \kw{bool} covers every case. Vx does not
yet have guards, or-patterns, range patterns, struct patterns, nested payload patterns, or
\kw{if let}.

\emph{Status:} structs, enums, tuples and tensors \st{tested}; pattern matching \st{partial}. Two
defects belong here. A struct that holds itself without a box makes the compiler overflow its
stack instead of giving an error (\#3786). And the coverage check for a match on a number runs only when the match gives a value. So a
match statement with no \kw{\_} arm can let a function
fall off its end (\#3785, Section~\ref{sec:secure-gaps}).

\subsection{Generics and traits}\label{sec:generics}

Generics are compiled by monomorphization: the compiler makes one copy of a generic function for
each set of type arguments. A type parameter can have trait bounds, written \kw{<T : Area>}. A
\kw{const} generic parameter can stand for a dimension, so a function can accept only
matrices whose inner dimensions agree (E3003 otherwise). Listing~\ref{lst:traits} shows a trait,
two impls and a bounded generic function.

\begin{lstlisting}[float=t,caption={A trait, two impls, and a generic function with a bound. It
prints \kw{12 32}.},label=lst:traits]
trait Area {
  fn area(self : &Self) -> f64;
}
struct Rect { w : f64, h : f64 }
struct Square { s : f64 }
impl Area for Rect {
  fn area(self : &Rect) -> f64 { return self.w * self.h; }
}
impl Area for Square {
  fn area(self : &Square) -> f64 { return self.s * self.s; }
}
fn twice<T : Area>(x : &T) -> f64 {
  return 2.0 * x.area();
}
\end{lstlisting}

Traits have default methods and associated types. The library defines the operator traits (\kw{Add}, \kw{Mul}, \kw{Neg} and the others) and the
comparison traits. It also defines \kw{Clone}, \kw{Copy}, \kw{Default}, \kw{From}, \kw{Hash} and
\kw{Drop}. The marker traits are \kw{Send}, \kw{Sync},
\kw{Owned} and \kw{HostOnly}. Dispatch is static. There is no \kw{dyn Trait}, no \kw{impl Trait}
and no \kw{where} clause except for the placement bound \kw{Reachable}
(Section~\ref{sec:placement}).

Two choices differ from Rust. First, a bound constrains the callers of a generic. But the compiler checks the body of a generic for each instantiation, as C++ checks a template. Second, when two
impls match one tensor receiver, the more specific shape pattern wins. For a generic struct, the
first impl in the file wins today (\#3787).

\emph{Status:} \st{tested}. Only the file that declares a struct or an enum can write a plain
\kw{impl} of it (E3066, \#3774). A macro that writes an impl counts as code of the file that calls
it. Another file can write an \kw{extern impl}, which is trusted code. Coherence still has gaps. A
generic struct can be \kw{Copy} when a field has the type of a parameter, with no bound on that
parameter. Such a struct copies what it holds, and a copied \kw{Vec} reads freed memory (\#3861). A type can be both \kw{Drop} and \kw{Copy}, and then
its drop never runs (\#3742).

\subsection{Closures and function values}\label{sec:closures}

A closure captures what its body uses. The capture mode follows from the body. A number that the body only reads is copied, and a local
that the body changes is borrowed mutably. A struct that the body reads is borrowed, and a value
that it moves is moved in. There is no \kw{move} keyword.

A closure type is
written \kw{|i64| -> i64}, and a function pointer type \kw{fn(i64) -> i64}. An \kw{unsafe fn} has a
type of its own, so it cannot pass where a safe function is expected. Listing~\ref{lst:closures}
shows the three capture modes.

\begin{lstlisting}[float=t,caption={Capture modes, from a program of the book that CI runs. It prints
\kw{double 8 add\_k 14}, then \kw{total 12}, then \kw{peek 3 count 3}.},label=lst:closures]
let double = |x : i64| x * 2;          // captures nothing
let k : i64 = 10;
let add_k = |x : i64| -> i64 { x + k };  // copies k
let mut total : i64 = 0;
let add = |x : i64| -> void { total += x; };  // borrows total mutably
add(5i64);
add(7i64);
let c = Counter { count : 3 };
let peek = || -> i64 { c.count };      // borrows c
\end{lstlisting}

A closure that copies a value keeps its copy. If the program changes the variable later and still
calls the closure, the warning W1034 says that the closure does not see the change.

\emph{Status:} \st{tested}. A closure whose body moves a value that it captures runs only once. A
second call is E4001. A copy, a return or a pass of such a closure is E4017 (\#3789).

\subsection{Ownership, borrowing and drops}\label{sec:ownership}

Vx takes its ownership model from Rust. A value moves unless its type is \kw{Copy}. The compiler tracks moves on each branch and on each pass of a loop. A read of a droppable field
moves it out of its struct.

A reference is shared (\kw{\&T}) or exclusive (\kw{\&mut T}). A value cannot have
an exclusive reference at the same time as any other reference. The call in
Listing~\ref{lst:borrow} is refused:
\begin{lstlisting}[caption={Two exclusive borrows of one variable in one call.},label=lst:borrow]
fn both(x : &mut i64, y : &mut i64) -> void { *x = *x + 1; *y = *y + 1; }
...
both(&mut n, &mut n);  // Error[E4004]: Cannot borrow 'n' because it is already
                       // borrowed as mutable.
\end{lstlisting}

A borrow ends at its last use, as with the non-lexical lifetimes of Rust. Fields of a struct borrow
separately, and a reborrow works. A tensor view is a borrow of its tensor. The borrow checker works
on the syntax tree of each function, one function at a time, so the compiler can run it in
parallel (P2). The compiler has about 7,800 lines of Rust for borrow, move and escape checks.

\paragraph{Regions without lifetime syntax.} Vx has no lifetime parameters. Inside a function, a
region is a block.

Across calls, the compiler needs to know where a returned reference may point. It
reads this from the body of the callee: for each function, a \emph{return-provenance
summary} records which parameters the result may point into. When the analysis cannot follow a case, the summary says that the result may point into
every reference parameter. Then the caller keeps all of its arguments borrowed. A programmer can state a sharper contract with a
\emph{borrow clause} in an \kw{ensures}. Listing~\ref{lst:clause} shows one.

\begin{lstlisting}[float=t,caption={A borrow clause. \kw{find} promises that its result does not
point into \kw{*key}. The compiler checks the promise against the body. So the caller can change
\kw{key} while it still holds \kw{r}. The program prints \kw{5 7}.},label=lst:clause]
fn find(v : &Vec<i64>, key : &i64) -> &i64
ensures !reaches(return, *key)
{
  for i in 0..v.len() {
    if *v.get_ref(i) == *key { return v.get_ref(i); }
  }
  return v.get_ref(0);
}
fn main() -> i32 {
  let mut v = Vec<i64>::new();
  v.push(3i64);
  v.push(5i64);
  let mut key = 5i64;
  let r = find(&v, &key);
  key = 7i64;              // allowed: r does not point into key
  println!("{} {}", *r, key);
  return 0;
}
\end{lstlisting}

Without the clause, the assignment to \kw{key} is refused with E4009. The summary of
\kw{find} cannot rule out that the result points into \kw{key}. The clauses are
\kw{points\_into(return, *p)}, \kw{reaches(return, *p)} and \kw{keeps(p)}, usually written with
\kw{!}. The compiler checks each clause against the body (E4016), and \kw{--print-borrow-clauses}
prints the clauses that it can prove.

\paragraph{Drops.} The compiler frees a value when its owner goes out of scope, with two policies.
A value that owns only memory, such as a tensor or a vector of numbers, is freed early. The
compiler frees it after its last use, or after the last use of a reference to it. A type with its own \kw{Drop} waits until the end of its
block, and values drop in the reverse order of their declarations.

A drop flag handles a value
that moves on one branch only. \kw{core::mem} has \kw{drop} and \kw{forget}. A panic does not unwind,
so drops do not run after a panic.

The documentation calls tensors that must not be dropped silently ``linear''. The implemented rule
is affine: an unused device buffer gets only the warning W1001, \kw{forget} exists, and
\kw{transfer} copies its input. The one must-use check is E6036, for an \kw{async} transfer token
that the program never waits for.

\emph{Status:} moves, borrows, borrow clauses and drops \st{tested}; linear values \st{designed}.
Issue \#1501 tracks the soundness of the borrow checker. It lists about 33 fixed bugs. Fourteen of
them came from 78 bug reports of the Rust compiler that the project translated to Vx. One drop defect is open, with two probe programs: a struct with a \kw{Drop} that holds a reference
can read freed memory (\#3784,
Section~\ref{sec:secure-gaps}).

\subsection{Errors and panics}\label{sec:errors}

Vx has no exceptions. \kw{Option} and \kw{Result} are library enums. The operator \kw{?} works on a \kw{Result}, and it returns the error from the function
unchanged. It does not convert the error, and it does not work on an \kw{Option} (E3063). Listing~\ref{lst:errors} uses \kw{?} and an enum of errors.

\begin{lstlisting}[float=t,caption={Errors as values, from a program of the book that CI runs. A
call with a sum over 99 gives \kw{TooLarge}, and \kw{report} prints
\kw{error: the sum 108 is too large}.},label=lst:errors]
enum AddError {
  NotADigit(i64),
  TooLarge(i64),
}
fn add(a : Result<i64, AddError>, b : Result<i64, AddError>) -> Result<i64, AddError> {
  let sum = a? + b?;
  if sum > 99 {
    return Result<i64, AddError>::Err(AddError::TooLarge(sum));
  }
  return Result<i64, AddError>::Ok(sum);
}
fn report_error(e : AddError) -> void {
  match e {
    AddError::NotADigit(c) => { println!("error: the character {} is not a digit", c); }
    AddError::TooLarge(n) => { println!("error: the sum {} is too large", n); }
  }
}
\end{lstlisting} \kw{panic(msg)} prints the message and the place of the call, then stops the
program. A parameter of type \kw{Caller} gets the place of the call, so a library function can
report the line of its caller. The library also has \kw{abort}, \kw{unreachable}, \kw{todo} and \kw{unimplemented}. A \kw{-> !} function whose body can finish is an error (E3028).

\emph{Status:} \st{tested}. There is no unwinding.

\subsection{Compile-time evaluation}\label{sec:comptime}

A \kw{comptime} block runs inside the compiler, and the compiler replaces it with its value.
Listing~\ref{lst:comptime} computes a factorial while it compiles. \kw{if comptime} removes the
branch that is not taken before the compiler checks it. A closure whose body is a \kw{comptime} block must fold at every call.

\begin{lstlisting}[float=t,caption={Compile-time evaluation. The block runs in the compiler, and
the program prints \kw{3628800}.},label=lst:comptime]
fn fact(n : i64) -> i64 {
  if n <= 1 { return 1; }
  return n * fact(n - 1);
}
fn main() -> i32 {
  let table = comptime { fact(10) };
  println!("{}", table);
  return 0;
}
\end{lstlisting}

The rule is \emph{must-fold}: a block that the compiler cannot finish is an error (E3033), and the
compiler never moves it to run time. The same holds for a block that writes a variable outside
itself. The evaluator has budgets: a loop stops after 1,000,000 iterations (E8005) and calls stop at
a depth of 256 (E8004). A false \kw{assert} in a block stops the compilation (E8002). Any other check that would stop
the program, such as a division by zero, gives E3033. An integer overflow in a
block has no value, so the block is refused.

Vx goes one step further than C++ \kw{constexpr} and Zig \kw{comptime}. A block may name a variable
that is known only at run time, if the result does not depend on it. The compiler then deletes the
block and keeps its value. The Must-Fold paper states the rule and proves, in Lean, that the
deletion is sound for a core calculus~\cite{kumar2026mustfold}.

A run-time twin of each block is the test oracle. The test runs the program with
the block, and again with the block as ordinary code. The outputs must agree. A generator made 57,950 such programs, and none folded
to a wrong value. A rerun of the first 33,493 at a later commit found none
either~\cite{kumar2026mustfold,kumar2026comptime}.

Macros are \kw{macro\_rules} definitions with \kw{expr}, \kw{ident} and \kw{ty} fragments. They
expand in a phase of their own, before any name resolves (P2), with a limit of 64 nested
expansions. A file can call its own macros and the macros of the files that it imports. A call that
can mean two macros is an error (\#3768). An error in code that a macro of another file wrote points
at the call (\#3771).

\emph{Status:} \st{tested}. The evaluator computes with 64-bit integers, and a test suite runs 13
programs through the run-time twin check in CI.

\subsection{Contracts}\label{sec:contracts}

A function can state a precondition (\kw{requires}) and a postcondition (\kw{ensures}, with
\kw{return} for the result). A loop can state an invariant, a struct can state an invariant, and an
\kw{assert} can state a fact. The compiler turns each contract into a query for the Z3
solver~\cite{demoura2008z3}. It must prove each \kw{ensures} at each return (E8001) and each
\kw{requires} at each call (E8006). At a call, it uses the \kw{ensures} of the callee. The
encoding is path-sensitive, and it models \kw{+}, \kw{-}, \kw{*} and negation with wrap-around at
the width of the type.

The wrap-around matters. The contract in Listing~\ref{lst:contract} looks true, but it is false for
the largest \kw{i32}, because \kw{x + 1} wraps. Z3 finds that case and the compiler refuses the
function. With the precondition \kw{requires x < 2147483647}, the compiler accepts it.
\begin{lstlisting}[caption={A postcondition that integer overflow breaks. Without a precondition,
the compiler refuses it with E8001.},label=lst:contract]
fn add_one(x : i32) -> i32
  ensures return > x
{
  return x + 1;
}
\end{lstlisting}

The solver runs as a separate process with a fixed resource limit. So one version of Z3 gives
the same answers each time, but Vx does not pin the version. When Z3 is missing, each contract fails (P3).

\emph{Status:} \kw{requires}, \kw{ensures}, loop invariants and \kw{assert} \st{tested}. There are 137 test programs that use \kw{requires} or \kw{ensures}. Of
these, 47 are in the directories of contract tests. Struct invariants are \st{partial}. Today they make the fields of the type private to its
module, but the compiler does not prove that each value meets the
invariant. A wrapper type \kw{Verified<T>} marks a result whose asserts were proved, but any code can
build one (\st{partial}). Division is not modelled.

\subsection{Unsafe code, foreign functions and inline MLIR}\label{sec:unsafe}

An \kw{unsafe} block allows what the compiler cannot check. Inside it, a program can read or
write through a raw pointer (E5002 outside \kw{unsafe}). It can call an \kw{unsafe fn} or a
foreign function (E5001), and it can make an uninitialized value (\kw{::uninit()}). An \kw{unsafe fn} still needs \kw{unsafe} blocks inside it.
The placement checks stay on in \kw{unsafe} code: a host read of device memory is E6003 there
too.

A raw pointer is \kw{*const T} or \kw{*mut T}, and it can name its memory space. A pointer can carry
\emph{pointer clauses} in a contract: \kw{valid(p, n)}, \kw{initialized(p, n)},
\kw{!overlaps(p, q, n)} and \kw{!frees(p)}. A safe call must prove the clauses of its callee, and an
\kw{unsafe} call trusts them. The clauses also tell LLVM that a pointer is not null and does not
alias another one. An \kw{assume} in \kw{unsafe} code states a fact without a proof.
Listing~\ref{lst:pointers} shows a function with two pointer clauses.

\begin{lstlisting}[float=t,caption={Pointer clauses. A safe caller of \kw{sum} must prove that
\kw{p} is valid and initialized for \kw{n} elements; an \kw{unsafe} caller trusts it, and
\kw{--audit} lists that trust.},label=lst:pointers]
fn sum(p : *const i64, n : i64) -> i64
requires valid(p, n)
requires initialized(p, n)
{
  let mut total = 0i64;
  for i in 0i64..n {
    total = total + unsafe { p[i] };
  }
  return total;
}
\end{lstlisting}

A foreign function is declared in an \kw{extern "C"} block. Its call is \kw{unsafe}, unless the
declaration marks it \kw{safe fn}, which is an unchecked promise. A \kw{safe} foreign function cannot
return a value of a type with an invariant (E8015). It also cannot take a mutable reference or
pointer to one, because C code does not keep the invariant. The \kw{ensures} of a foreign function is trusted. A
\kw{pub fn} is exported with the C calling convention, so C code can call Vx.

An \kw{mlir!} block embeds MLIR operations. It declares its inputs, its results, the dialects that it
uses and what it clobbers. The compiler checks the list of dialects (E8010) and parses the
operations. Outside \kw{unsafe}, a block must be one that the compiler can verify as safe (E8016).
Such a block uses only the \kw{arith} and \kw{math} dialects. It takes and gives only numbers or
vectors of numbers, and it writes no variable of the program. Each of its operations is on a list of operations that give a defined result for every input, such as
\kw{arith.addi} with no overflow flag. A load from an address, an integer division and an add with
an overflow flag are not on the list, so they need \kw{unsafe}. Only the file of a type with
an invariant can make a value of that type in an \kw{mlir!} block (E8014). This rule holds in
\kw{unsafe} code too.

\emph{Status:} \kw{unsafe}, raw pointers, pointer clauses, \kw{extern} and \kw{mlir!}
\st{tested}. A method call through a raw pointer also needs \kw{unsafe} now (E5002, \#3798). The
compiler does not check what the operations of an \kw{mlir!} block in \kw{unsafe} code do. The check
for E8016 has a gap: it reads only the first operation of each line. So a block with two operations
on one line can load from an address with no \kw{unsafe} (\#3864). \kw{--audit} still
lists such a block.
Section~\ref{sec:trusted} returns to the trusted code.

\subsection{Capabilities}\label{sec:capabilities}

A capability is a value that gives the right to do one thing. Code that does not hold the value
cannot do the thing. Vx makes capabilities from the rules that it has for a struct with an invariant
(Section~\ref{sec:contracts}). Three rules hold for such a type:
\begin{itemize}
\item \emph{No forging.} Only the file that declares the type can build a value of it (E8014). An
  \kw{mlir!} block in another file cannot make one (E8014), and a \kw{safe} foreign function cannot
  return one (E8015).
\item \emph{No copying.} A struct moves (E4001). Only the file that declares a type can write a plain
  \kw{impl} of it (E3066), so another file cannot make it \kw{Copy}. Another file can write an
  \kw{extern impl}, which is trusted code (Section~\ref{sec:trusted}).
\item \emph{No reference after the owner.} A reference to a capability cannot outlive the value that
  it borrows (E4005), apart from the gap of \#3784. Code can keep a lent reference while the owner
  holds the value.
\end{itemize}
The invariant can be \kw{invariant true}, which states no fact. It only makes the type one that its
own file alone can build.

The standard library uses these rules for the operating system. The module \kw{std::root} declares
\kw{Root}, which holds two rights. \kw{Files} is the right to open files, and \kw{Network} is the
right to open connections and to bind addresses. A program gets its \kw{Root} as the parameter of
\kw{main}, and it gives each right to the code that needs it (Listing~\ref{lst:caps}).
\kw{File::open}, \kw{TcpStream::connect}, \kw{TcpListener::bind} and \kw{UdpSocket::bind} take a
reference to a right as their first parameter. So a library that gets no \kw{\&Files} cannot call
\kw{File::open} in safe code. A \kw{main} with no parameter is still correct, and a program that needs no right
keeps that form.

\begin{lstlisting}[float=t,caption={Capabilities. The parser gets no right, so it cannot call
\kw{File::open}. The cache gets the file right only during one call.},label=lst:caps]
import std::fs;
import std::string;

// The parser gets no right, so it cannot call File::open.
fn parse_name(request : &String) -> String {
  return String::new(); // reads the name from the request
}

// The cache gets the file right only during the call.
fn open_cached(files : &Files, name : &String) -> Result<File, FileError> {
  return File::open(files, name, OpenMode::Read());
}

fn main(root : Root) -> i32 {
  let request = String::new();
  let name = parse_name(&request);
  match open_cached(&root.files, &name) {
    Result<File, FileError>::Ok(_file) => { return 0; }
    Result<File, FileError>::Err(_e) => { return 1; }
  }
}
\end{lstlisting}

\emph{Status:} the three rules and the rights \kw{Files} and \kw{Network} \st{tested}. For each
right, a test tries to copy it, to build it, and to open a file or a socket without it. Each try
fails. A right to one
folder is \st{designed} (\#3750). Rights for processes, memory maps and native code have no design
yet. Trusted
code can still reach the operating system with no right: an \kw{unsafe} block can call
\kw{File::open\_c\_str} or a C function. Gaps let safe code copy a right, or build a \kw{File} or a socket from a raw pointer
(Table~\ref{tab:gaps}). Section~\ref{sec:authority} says what the
rights give and where they stop.

\section{Scalable}\label{sec:scalable}

A program scales when it can use more machines, more cores and more GPU threads without a new
program text. A compiler scales when a larger program does not take much longer to check. This
section covers both. It starts with the declarations of machines, then follows a value from a host to a device and back.
Then it describes three levels of parallel code: tasks on CPU cores, \kw{across} loops on GPU
threads, and work on other machines.

\subsection{Machines as declarations}\label{sec:machines}

A machine file is a Vx file that contains only \kw{Memory} and \kw{Topology} declarations. It
describes one part: its memory spaces, their capacities and bandwidths, the links between them,
and the levels of threads that run code. Listing~\ref{lst:machine} shows part of the file for an
NVIDIA H100.

\begin{lstlisting}[float=t,caption={Part of \kw{fleet/h100-sxm.vx}, with the comments shortened.
Most figures have a source comment. Some on-chip figures have none; the header of the file calls
them unverified.},label=lst:machine]
// spec: NVIDIA H100 product page, SXM column (verified 2026-08-03)
Memory HBM {
  capacity: 80 GiB, bandwidth: 3.35 TB/s, managed: explicit, alignment: 16 B, scope: device
}
// spec: NVIDIA Hopper Architecture Whitepaper -- 50 MB L2 cache.
Memory L2 {
  within: Memory::HBM, capacity: 50 MiB, bandwidth: 12 TB/s, managed: cached, scope: device
}
// replicas: measured: probe_edges reported SMs=132 on two rented H100 SXM boxes
Memory SMEM {
  within: Memory::L2, capacity: 228 KiB, bandwidth: 128 B/cyc, clock: 1.98 GHz,
  granule: 1 KiB, managed: explicit, alignment: 16 B, scope: sm, replicas: 132
}
// spec: PCIe Gen5 x16 = 63.0 GB/s per direction. UNVERIFIED on this SKU.
Topology Device {
  arch: nvptx64, chip: sm_90a,
  levels: [blocks(max_lanes: 1024, memory: cta), subgroups(lanes: 32), lanes(memory: thread)],
  memory: Memory::HBM, visible: [Memory::HBM, Memory::L2, Memory::SMEM, Memory::REG],
  transfer Memory::CPU_DRAM -> Memory::HBM : 63 GB/s,
  transfer Memory::HBM -> Memory::CPU_DRAM : 63 GB/s,
  transfer Memory::HBM -> Memory::L2,
  transfer Memory::L2 -> Memory::SMEM copy_engine
}
\end{lstlisting}

A \kw{Memory} declaration gives a capacity, a bandwidth and the space that contains it (\kw{within}).
It says whether the hardware manages the space (\kw{cached}) or the program does (\kw{explicit}).
It also gives the scope of the space: device, multiprocessor, block or thread. Last, it gives the
number of copies (\kw{replicas}), the allocation unit (\kw{granule}) and the alignment.

A \kw{Topology} gives the architecture, the default space and the spaces
that its code can see. It also gives the element types that it supports, its levels of threads, and
its transfer edges. An edge has at most one cost source (E6013 for a second one). The source is a link
rate, a unitless weight, or a bandwidth that the compiler derives from \kw{within}.

The compiler checks the file itself. A cycle of \kw{within} links (E6006), a space larger than
its parent (E6007) and a figure that is not positive (E6008) are errors.

The compiler ships 16 machine files. Ten describe accelerators. Six are NVIDIA GPUs: A100 40 and 80~GB,
H100 SXM and PCIe, H200 and B200. The other four are an AMD MI300X, an Intel Arc A770, an Apple M4
with its unified memory and neural engine, and an Arm Cortex-M7 microcontroller. The others describe nodes with two
and eight GPUs, and three x86 hosts.

A program is
compiled against one of them with \kw{--machine}, and \kw{--host} names the host. The program text does
not change. In CI, 249 test files compile against a machine file.

\emph{Status:} \st{tested} without hardware: every check runs on the compiling machine. The
provenance comments are a convention that no tool checks (P4). A machine file cannot import another,
so the node files repeat the declarations of their GPUs (\#3350).

\subsection{Placement in types}\label{sec:placement}

The type of a tensor names its memory space: \kw{Tensor<f32, [4, 4], Memory::HBM>}. A tensor with no
space lives in host memory. A raw pointer can name its space too, and a pointer into one space does
not pass where a pointer into another is declared (E6029). A function can be generic over the space. The bound \kw{Reachable<A, B>} says that a declared route takes data from topology \kw{A} to
topology \kw{B}.

\kw{spawn on(Topology::Device) \{ \ldots \}} runs a block on a device. The compiler outlines the block into a kernel. It checks that the device can address every value
that the block reads (E6003, as in Section~\ref{sec:intro}). A plain function belongs to the host, so a call to it from a device
block is an error (E6001). A channel end or a lock in a device block is an error too (E6038),
because the device cannot wait on host memory. The meaning is sequential: the statement after a
\kw{spawn on} sees the block as finished.

\emph{Status:} \st{tested}. Only device 0 of a declared GPU runs code today (E6042 for the others). Two
open issues are gaps in the check. A \kw{spawn on} region that ends in a number now gets a GPU kernel
(\#3715). Other values at the end of such a region stay on the host (\#1809). A kernel can call a CPU-only function through a method (\#3638).

\subsection{Transfers and routes}\label{sec:transfers}

\kw{transfer(x, Memory::M)} gives a copy of \kw{x} whose type names \kw{M}. The compiler finds a route
over the declared edges, and a missing route is an error (E6002). It splits a route with several
hops into one checked transfer for each hop.

To choose a route, the compiler first looks for routes whose edges all have a price per byte,
from a bandwidth or a link rate. It takes the cheapest of them. If no route has prices on all of its edges, it uses the unitless weights. The price is per byte,
so the route does not depend on the size of the tensor. A test compares each chosen route with a
Bellman--Ford search over every machine file.

\kw{transfer\_async} gives a token, and \kw{.wait()} on the token gives the placed tensor. A token that
the program drops without a wait is an error (E6036). The compiler does not choose placements: the
programmer writes each transfer, and the cost model only explains the cost.

A machine can also say in Vx how it moves bytes across an edge. It writes a body for
\kw{impl Transfer<Memory::A, Memory::B> for Topology::X}. The body may use only low-level primitives
(\kw{raw::load}, \kw{raw::store}, \kw{raw::barrier}, \kw{raw::async\_copy}). The compiler proves the
bounds of each primitive with Z3 and refuses a barrier under a condition.

\emph{Status:} transfers and routes \st{tested}; user transfer bodies \st{partial}, because the
compiler emits them only for edges into multiprocessor memory. The model has no per-transfer latency, so the predicted time of a small copy is far too
low. On an A100 host link, it is about 99\% too low at 4~KiB~\cite{kumar2026machinemodel}. The model also has no contention between flows.

\subsection{Capacity}\label{sec:capacity}

The compiler adds the bytes that a program places in each space that has a capacity, and it refuses
a program that does not fit. E6009 is a single tensor that is larger than its space. Against the
H100 file, the admission program below gives:
\begin{lstlisting}[language={},basicstyle=\ttfamily\scriptsize]
Error[E6009] at admit.vx:87:3: transferred tensor needs 141104775168 bytes but memory space
'HBM' has capacity 85899345920 bytes
\end{lstlisting}

E6010 is a set of tiles that are alive at the same time and do not fit together. The quantity is a peak over scopes, not a sum over the whole program. A tile lives until its
last reader, and the end of an \kw{if} arm or of a loop body frees what it placed. Granules round each tile up, and a space with
\kw{replicas} is checked for one copy.

Across calls, each function exports a summary of its peak for each space. A fold over
the call graph adds the peak of a callee to what is alive at the call (E6027). Tasks that run at the same time add up. A recursion that places tiles in a space with a
capacity is refused (E6028).

For the same program and machine, the working set of three tiles gives:
\begin{lstlisting}[language={},basicstyle=\ttfamily\scriptsize]
Error[E6010] at admit.vx:83:4: the working set placed in memory space 'HBM' (3 tiles) sums to
142514061312 bytes, over its 85899345920 byte capacity; place fewer/smaller tiles or declare it
`overcommit`
\end{lstlisting}

\paragraph{One program, many machines.} The compiler ships with an admission program,
\kw{fleet/admit.vx}. It serves Llama-3.1-70B in 16-bit floating point, with a context of 4,096 tokens and a batch
of one. It computes the sizes of the weights, the KV cache and the activations from the model's
shape parameters, with \kw{const} generic arithmetic. The file never changes between
machines. Table~\ref{tab:admit} shows its verdict against each machine file that describes a device,
from a build of the main branch. The weights alone need 141,104,775,168 bytes (131.4~GiB), so only the
three parts with more memory admit the program.

\begin{table}[t]
\caption{The verdict of one unchanged program (\kw{fleet/admit.vx}: Llama-3.1-70B, 16-bit, 4k
context, batch 1) against 15 machine files, at \kw{719726333}. The capacity is that of the space that
the program fills.}\label{tab:admit}
\small
\begin{tabular}{@{}llll@{}}
\toprule
Machine file & Capacity & Verdict & Codes \\
\midrule
\kw{b200} & 183,359 MiB & admitted & --- \\
\kw{h200} & 141 GiB & admitted & --- \\
\kw{mi300x} & 192 GiB & admitted & --- \\
\kw{h100-sxm}, \kw{h100-pcie}, \kw{a100-80} & 80 GiB & refused & E6009, E6010 \\
\kw{node-2gpu-a100}, \kw{node-2gpu-h100}, \kw{node-8gpu} & 80 GiB (one GPU) & refused & E6009, E6010 \\
\kw{a100-40} & 40 GiB & refused & E6009, E6010 \\
\kw{m4-uma} & 32 GiB & refused & E6009, E6010 \\
\kw{arc-a770} & 16 GiB & refused & E6009, E6010 \\
\kw{xeon-8275cl}, \kw{xeon-e5-2666v3} & 96, 30 GiB & refused & E6009, E6010 \\
\kw{cortex-m7} & 512 KiB & refused & 3 $\times$ E6009, E6010 \\
\bottomrule
\end{tabular}
\end{table}

The node files refuse the program because it does not split the model across GPUs: it uses one GPU of
the node. A version with tensor parallelism divides each size by the number of GPUs, and the study of
LLM serving uses such versions~\cite{kumar2026admission}.

\kw{--diagnostics-json} writes the verdict as a record for other tools. Each capacity error carries
the space, the required and available bytes, and the margin. Each route carries its edges, its bytes,
a derived time, and the traffic in each space. A study of LLM serving used these records. Before any GPU rental, it predicted which
configurations fit on which GPUs. All 24 launches on an H100 and an A100 agreed with the
predictions~\cite{kumar2026admission}.

\emph{Status:} \st{tested}. A call through a function value is not counted (\#3632), and a dynamic
shape skips the check with the warning W1029. The capacity is the capacity of the device, not the
budget that a serving engine leaves.

\subsection{Seam checks}\label{sec:seams}

A seam is one hop of a transfer. With \kw{--verify-seams}, the compiler builds a bit-vector query for
each hop and runs Z3. A transfer that synchronizes keeps the value of a buffer. A transfer that the program or the machine file marks \kw{relaxed} makes the value unknown. A read of
the value after the hop is an error (E6004), and the Z3 model is the counterexample. When Z3 is missing, the check fails (E6024), unless
\kw{VX\_ALLOW\_UNVERIFIED=1} turns the errors into the warning W1031.

\emph{Status:} \st{tested}, and off by default. Only three test files turn it on. The answer depends only on whether a hop is relaxed, so the query decides what a syntactic
check would decide. The value part of the contract compares a value with itself (\#662). There is no formal model of
seams, transfers or capacity yet (\#1209).

\subsection{Tasks on CPU cores}\label{sec:tasks}

Vx has structured concurrency. A \kw{tasks} block starts tasks with \kw{spawn}, and the block ends only
after each of its tasks ends. So a task can borrow the locals of the code around it, as in
Listing~\ref{lst:tasks}.

\begin{lstlisting}[float=t,caption={Two tasks each add half of a vector that they borrow from
\kw{main}. The program prints \kw{total 5050}.},label=lst:tasks]
fn sum_range(v : &Vec<i64>, from : i32, to : i32) -> i64 {
  let mut sum : i64 = 0;
  for i in from..to { sum = sum + *v.get_ref(i); }
  return sum;
}
fn main() -> i32 {
  let mut v = Vec<i64>::new();
  for i in 1..101 { v.push(i as i64); }
  let total = tasks {
    let low = spawn { sum_range(&v, 0, 50) };
    let high = spawn { sum_range(&v, 50, 100) };
    low.join() + high.join()
  };
  println!("total {}", total);
  return 0;
}
\end{lstlisting}

\kw{h.join()} gives the value of a task. A \kw{spawn} outside a \kw{tasks} block is an error (E3056),
and a handle cannot leave its block. A task whose body gives a \kw{Result} can fail. Its failure
leaves the block, and the other tasks get a cancel. When nothing uses the value of a \kw{tasks} block
that can fail, the compiler gives the warning W1038, and the block drops the error.

There is no \kw{async} and no \kw{await}: a wait is
a normal call, and the same program runs on either runtime. The default runtime gives each task an
operating-system thread. A second runtime runs tasks as fibers on a pool of worker threads, from one first-in,
first-out queue. For tests, a third runtime runs all tasks on one thread, in an order that a seed
chooses.

The borrow checker treats each task as a borrow that lasts from its \kw{spawn} to its join. So two
tasks that write one variable are refused:
\begin{lstlisting}[language={},basicstyle=\ttfamily\scriptsize]
Error[E4002] at race.vx:8:7: Cannot access 'x' because it is mutably borrowed. The task
from the `spawn` at line 4 uses it at line 5, and it is not joined yet.
\end{lstlisting}
A task that moves or writes a value needs \kw{Send}, and a task that reads a shared value in place
needs \kw{Sync}. The compiler derives both from the fields of a type.

The library has bounded channels, and also one-shot, broadcast and watch channels and a work
queue. The functions \kw{select2} to \kw{select6} wait on sources of different types. A source can be
the receiver of a channel of each kind, a socket, a \kw{Semaphore}, an \kw{Interval} or a
\kw{Deadline}. Each call gives the index of the first source that is ready. The caller then takes
the value, so a source that loses keeps its value. A \kw{Mutex}
gives its lock in the order of the requests. The library also has \kw{Semaphore}, \kw{Arc}, and
atomics with the memory orders of C++. Vx compiles each
\kw{Relaxed} read as an \kw{Acquire} read, to rule out values from thin air; a flag keeps relaxed
reads relaxed.

When every task waits and nothing can wake one, a watcher thread stops the program. It
prints each wait, the line that made each channel or lock, and the cycle of waits. For a
\kw{select}, it names each channel and semaphore that the call waits on (\#3816). In the
following run, each of two tasks waits to receive from the other before it sends:
\begin{lstlisting}[language={},basicstyle=\ttfamily\scriptsize]
vx: deadlock: each task of the program waits, and nothing can wake one. The program stops.
  ...
  the waits form a cycle:
    the task from the `spawn` at dl.vx:26 waits in `recv` at dl.vx:10 on channel 1, made at
      dl.vx:21, which only the task from the `spawn` at dl.vx:29 can send on
    the task from the `spawn` at dl.vx:29 waits in `recv` at dl.vx:10 on channel 2, made at
      dl.vx:22, which only the task from the `spawn` at dl.vx:26 can send on
  the channels of these waits:
    channel 1, made at dl.vx:21: 1 wait in `recv` and 0 in `send`. 1 `Sender` is not dropped.
    channel 2, made at dl.vx:22: 1 wait in `recv` and 0 in `send`. 1 `Sender` is not dropped.
\end{lstlisting}

\emph{Status:} tasks, channels, \kw{Mutex}, \kw{Semaphore}, \kw{Arc} and atomics \st{tested} on
x86-64 in CI; \kw{select} over sources of different types \st{tested} (\#3807, \#3814, \#3815). The deadlock watcher is
\st{partial}. It reports only when every thread waits. On a 0.0.4 build, it did not report a
deadlock next to a task that spins. The borrow check of a task holds whole variables, not fields: two tasks that write two
different fields of one struct are refused. Section~\ref{sec:concurrency} describes the proofs and
model checks of this design.

\subsection{Loops on GPU threads}\label{sec:across}

A GPU kernel in Vx is a loop nest. A word after the range of a \kw{for} loop says which level of the GPU runs the passes of the loop. The words are \kw{across blocks}, \kw{across subgroups}, \kw{across lanes} or
\kw{across lanes(n)}. The levels come from the machine file. Listing~\ref{lst:across} doubles each
element of a matrix in GPU memory and adds the number of its row.

\begin{lstlisting}[float=t,caption={A GPU kernel as a loop nest. Each pass of the outer loop runs in a
block of its own, and the inner loop runs on 256 lanes. With the H100 file, on a Mac with no NVIDIA GPU, the CPU copy prints \kw{corner 2, last row 65};
without the \kw{across} words the program prints the same.},label=lst:across]
fn main() -> i32 {
  let mut grid = transfer(Tensor<f32, [64, 300]>::fill(1.0), Memory::HBM);
  spawn on(Topology::Device) {
    for y in 0..64 across blocks {
      for x in 0..300 across lanes(256) {
        grid[y][x] = grid[y][x] * 2.0 + (y as f32);
      }
    }
  }
  let back = transfer(grid, Memory::CPU_DRAM);
  println!("corner {}, last row {}", back[0][0], back[63][299]);
  return 0;
}
\end{lstlisting}

The meaning of the program does not depend on \kw{across}: delete each \kw{across}, and the loops
run in order with the same result. This holds for kernels that do not call a \kw{lanes::} routine,
which needs its \kw{across} loops. For this to hold, the compiler must show that no two passes touch an
element that one of them writes. For each tensor that a loop nest writes, every access must use an index that fixes the loop
variables.

The compiler checks this on affine index forms, with a mixed-radix separation rule. It accepts constant gaps (\kw{2*j} and \kw{2*j+1}) and division and remainder by a factor of
the loop end. It also accepts nested digits such as \kw{(l \% 32) / 4}. It refuses a write that another pass may also write, and a read of an element that another pass
writes. It also refuses a write to a variable from outside the loop. If the kernel of Listing~\ref{lst:across} writes
\kw{grid[0][x]} instead of \kw{grid[y][x]}, every block writes row 0, and the compiler refuses it:
\begin{lstlisting}[language={},basicstyle=\ttfamily\scriptsize]
Error[E6032] at race.vx:6:9: two passes of this block loop may write the same element of
`grid`: an index of the write must fix `y`
\end{lstlisting}

Two wrapper types add one fact each. The type \kw{Distinct} says that the elements of an index tensor differ. The type
\kw{DisjointRanges} says that a table of ranges does not overlap.

The end of each lane loop is a barrier for the lanes of a block, so every lane must reach it. The
checker requires that each condition around a lane loop be the same for every lane of the block.
Code between lane loops that writes memory runs on lane 0 only, and a barrier follows it.

The compiler lowers a nest to grid-stride loops in the MLIR \kw{gpu} dialect. Then it
lowers them to PTX for NVIDIA, to the Metal Shading Language for Apple GPUs, or to SPIR-V for Intel
GPUs. Each kernel also has a CPU copy, where a lane index reads as 0 and a block count as 1. So a test
can compare the GPU result with the CPU result.

The Erasable Parallelism paper states the race rule with a proof sketch. It applies the rule to 38
kernels from llm.c, PolyBench/GPU and Rodinia. The checker accepts 27 and refuses 11, and three
of the refusals are real races~\cite{kumar2026erasable}.

\emph{Status:} the checks \st{tested} without hardware, as static checks. Kernels on NVIDIA \st{tested by hand}, with A100, A6000, H100 and H200 runs recorded in commits
and issues. CI checks the PTX or LLVM text of 93 programs, but it has no GPU. Kernels on Apple GPUs \st{tested by hand}: 29 programs need
macOS, and CI has no macOS test job. The race check has no mechanized proof. The older of the two code
generators drops some barriers (\#3062), and the default one can fall back to it.

\subsection{Tensors and machine learning}\label{sec:tensors}

A tensor type has an element type, a shape and a space. The element types include \kw{bf16}, the
eight-bit floating-point formats and four-bit integers. A dimension can be \kw{?}, for a size that is
known only at run time. The library has element-wise operations, reductions, \kw{@} for matrix
products, and views.

Automatic differentiation is a built-in form: \kw{grad(loss, 3.0)} gives the
derivative, and \kw{vjp} and \kw{jvp} give vector--Jacobian and Jacobian--vector products. The
compiler hands the work to Enzyme~\cite{moses2020enzyme}, which runs as an LLVM plugin. The module \kw{std::nn} has 49 functions for the decoder of a transformer, with \kw{bf16}, eight-bit and four-bit weights. The module \kw{std::collective} calls NCCL for
all-reduce and all-gather.

\emph{Status:} tensors and \kw{@} \st{tested}; automatic differentiation \st{tested} for scalar
functions only. The checker does not yet refuse a tensor argument, which then fails inside Enzyme. The module \kw{std::nn} is \st{tested by hand} on an Apple GPU, with CPU tests in CI. Collectives are \st{partial},
tested with one rank on a CPU. The inner dimension of \kw{@} on two tensors of shape \kw{[?, ?]} is not
checked. So such a product can read past the end of a tensor, also with \kw{-fbounds-checks} (\#3792, Table~\ref{tab:gaps}).

\subsection{Compiling large programs}\label{sec:frontend}

The design rules of P2 let the compiler check function bodies in parallel. Every symbol and type has a 256-bit global identifier (GID). The
compiler computes it from the module path and the name, so it does not depend on which thread
finds the symbol first. A generic instantiation gets a number from a serial step, in a fixed
order. With \kw{-j N}, the compiler runs its frontend in phases. Most phases are a parallel map over modules or functions that
reads a frozen table. The registry freeze, the declaration check and the merge of new generic
instantiations are serial steps, and the merge works in a fixed order.

The output does not depend on the number of threads. CI compiles programs at 1 and 8 threads and with a
sequential schedule, and the generated MLIR must be the same bytes. A second gate compares 1 and 4
threads on a program of 1,000 modules and 16,000 functions. An August build of the frontend compiled to MLIR 5.9 times faster on 48 Xeon cores, with
the same output at every thread count~\cite{kumar2026frontend}. A September run on another
48-core machine gave 3.9 to 4.5 times. On a whole compile, the serial MLIR back end takes most of
the time.

\emph{Status:} \st{tested}, but opt-in. A plain \kw{vxc file.vx} runs a sequential driver, whose output
differs in the order of functions and which lowers only the code that \kw{main} reaches. The parallel path takes no lock in Vx code. But each phase is a fork and join of a
thread pool, and the merge between phases is serial. Six flags do not work with \kw{-j} yet, among them \kw{--verify-seams},
\kw{--diagnostics-json} and \kw{--print-borrow-clauses}.

\subsection{Work on other machines}\label{sec:remote}

A topology with \kw{offload: true} may run on a worker process on another machine. A manifest maps
topologies to workers when the program starts, and a wire protocol carries the buffers. A host stops
with an error when it loses a worker: the host checks the link and stops after 30 seconds with no
answer. A \kw{Memory} with \kw{node: N} is a NUMA domain of the host. The runtime binds a placement to
its node and pins the thread that reads it. On a two-socket \kw{c5.metal}, the predicted ratio of
remote to local cost was 2.26 and the measured ratio 2.35.

\emph{Status:} \st{tested} on one machine, with workers on the loopback interface; one run on two
machines is recorded. The protocol uses plain TCP with no authentication
(Section~\ref{sec:toolchain}). NUMA binding is \st{tested by hand}.

\subsection{Backends}\label{sec:backends}

Table~\ref{tab:backends} lists what runs on each target. Two points need care. First, the CPU backend does not use an in-process JIT. The compiler runs
\kw{mlir-translate}, \kw{opt}, \kw{llc} and \kw{clang} as child processes, then runs the program.

Second, the compiler does not generate code for
the Apple neural engine. The runtime has four prebuilt CoreML models, and it chooses one by three patterns of a
region. Every other region on the neural engine runs on the CPU. A region that the compiler tags as a $4 \times 4$ matrix product gets a model that CoreML plans
on the CPU. The square 16-bit products of size 512 and 1024 get models that CoreML plans on the
neural engine.

One route chooses a model by the shapes of the buffers alone. It gives wrong results for each
kernel that it matches, also for an affine one, and it writes the result into an input (\#3793).

\begin{table}[t]
\caption{Backends at \kw{719726333}, and how each is tested.}\label{tab:backends}
\small
\begin{tabularx}{\linewidth}{@{}lYl@{}}
\toprule
Target & Path & Status \\
\midrule
CPU (x86-64, AArch64) & MLIR $\to$ LLVM IR $\to$ object file, through child processes &
  \st{tested} (x86-64 in CI) \\
NVIDIA GPU & MLIR \kw{gpu} $\to$ NVVM $\to$ PTX; CUDA driver; cuBLAS for tagged products &
  \st{tested by hand} \\
Apple GPU & MLIR \kw{gpu} $\to$ Metal Shading Language & \st{tested by hand} (macOS) \\
Apple neural engine & four prebuilt CoreML models, chosen by pattern; else the CPU &
  \st{partial} \\
Intel GPU & SPIR-V, Level Zero, oneMKL; \kw{across} loops, lane routines, shared tiles and
  atomics; no CI & \st{partial} (one card run) \\
AMD GPU, Cortex-M7 & machine files only; no code generator & \st{designed} \\
Remote worker & wire protocol over TCP & \st{tested} (loopback) \\
\bottomrule
\end{tabularx}
\end{table}

A device of a declared machine does not have to be present. On a computer with no GPU, the runtime runs the CPU copy of each kernel,
so the result is the same. A build with CUDA prints that it found no device. Some CPU copies are
stubs that stop the program, such as the copy of an \kw{mlir!} block for a thread group. The
compiler can produce PTX for a declared H100 on a laptop, but the host object always gets the triple
of the compiling machine.

\section{Maintainable}\label{sec:maintainable}

A program is maintainable when a person can change one part of it without first reading all of
it. The tools must also say what a change broke. This section describes the parts of Vx that serve that goal: rules for local
reasoning, modules and interfaces, diagnostics, determinism and tools. It ends with the practices
that the project uses to keep its own claims true.

\subsection{Local reasoning}\label{sec:local}

The rules of P2 also serve a reader. A call names one function, because there is no overloading. A
value changes type only where the program writes \kw{as}, because there are no implicit conversions.
A function's signature gives the types of every parameter and of the result, and its contract states
what it requires and ensures.

A transfer between memory spaces is a call in the source, not a copy
that a runtime makes. And a \kw{comptime} block either folds or fails. So a reader does not have to ask whether a block
ran in the compiler or at run time.

Vx does not reach this goal everywhere. A return-provenance summary comes from the body of the
callee (Section~\ref{sec:ownership}), so a change to a body can change which callers compile. A borrow clause in the signature fixes the contract. The flag \kw{--print-borrow-clauses} shows the
clauses that the compiler can prove, so a library author can write them down.

\subsection{Modules, interfaces and versions}\label{sec:modules}

A file is a module, and \kw{import} brings in what another module declares. The compiler looks for an import in the standard library first, then in the directory
of the file that it compiles. It never looks in the current directory, so the result of a build
does not depend on where it runs. \kw{--package
name=dir} names a library directory. With \kw{-j}, modules load in waves, and the compiler parses each wave in parallel.

A library can be compiled to an interface file with \kw{vxc --emit-interface lib.vx -o lib.vxlib}.
The file holds the frozen table of signatures, layouts and impls. It also holds the flat-HIR bodies of non-generic functions, and it
leaves out generic bodies. A program then links the interface without the source. The reader of the format
checks the bounds of every read and fails closed.

\kw{vx-api diff OLD NEW} compares two interfaces and prints the smallest version change that the
differences need, in the terms of semantic versioning. A removed function and a new parameter type each need a major version. So do a result that now
borrows from another parameter, and a function that becomes \kw{unsafe}. A new function or struct needs a minor version. For structs and enums, an added field or an added variant also needs a major
version. The reason is that a client may build the value or match on it.

Seventeen tests check the tool. In the following example, version 2 of a small library changes the
type of a parameter of \kw{scale} and adds a function \kw{origin}:
\begin{lstlisting}[language={},basicstyle=\ttfamily\scriptsize]
$ vx-api diff v1/shapes.vxlib v2/shapes.vxlib
major  function `scale` changed from `fn(&Point, i64) -> Point` to `fn(&Point, i32) -> Point`
minor  function `origin` was added
not compared: traits, enums whose variants carry data, constants, and the bounds of generic parameters
bump: major
\end{lstlisting}
The tool states what it did not compare, so a reader knows the limits of its answer.

\emph{Status:} modules, imports and interfaces \st{tested}; \kw{vx-api} \st{tested}. The tool states
four blind spots: traits, enums with data, constants and generic bounds. It has others that it does
not state. It does not compare contracts or visibility, and it skips the signatures that the
interface format skips. So the change that it reports can be too small.

Visibility is minimal today: \kw{pub fn}
exports a symbol from an object file, and every top-level item is visible to every importer. A design
with four levels (private, package, workspace, public) and a version rule for each level is
\st{designed}, and a companion paper develops it. A package manager does not exist yet (\#1275).

\subsection{Diagnostics}\label{sec:diagnostics}

The compiler defines 196 diagnostic codes: 167 errors and 29 warnings. The first digit of a code
names the stage that raises it (Table~\ref{tab:codes}).

\begin{table}[t]
\caption{The diagnostic codes of \kw{vxc} by group, from the generated diagnostic index.}\label{tab:codes}
\small
\begin{tabularx}{\linewidth}{@{}llrY@{}}
\toprule
Group & Range & Codes & Example \\
\midrule
Warnings & W1001--W1038 & 29 & W1022: a \kw{transfer} into the space that the value is already in \\
Parser & E1001--E1013 & 13 & defined, not yet emitted (\#3227) \\
Names & E2001--E2007 & 7 & E2001: an undefined variable \\
Types & E3001--E3066 & 65 & E3033: a \kw{comptime} block that cannot fold \\
Ownership and borrows & E4001--E4017 & 17 & E4004: a second exclusive borrow \\
Safety & E5001--E5003 & 3 & E5002: a raw-pointer access outside \kw{unsafe} \\
Placement and machines & E6001--E6042 & 42 & E6003: a value that a device cannot address \\
Tensors and math & E7001--E7004 & 4 & E7001: shapes that do not agree for \kw{@} \\
Contracts & E8001--E8016 & 16 & E8001: a postcondition that the solver cannot prove \\
\bottomrule
\end{tabularx}
\end{table}

Each message names the place, the fact that failed, and often the fix, as E6003 and E4002 above show. The project's web site has one page
for each code, with a program that triggers it where the test suite has one. A suggestion for a misspelled name comes from the edit distance to the names in scope. The flag
\kw{--diagnostics-json} writes the same facts as a record for tools (Section~\ref{sec:capacity}).
The flag \kw{--audit} lists the trusted facts (Section~\ref{sec:trusted}).

\emph{Status:} \st{tested}. 22 of the 196 codes are never emitted, and some errors have no code
(\#3227). The JSON record does not name the compiler version that wrote it.

\subsection{Determinism and reproducibility}\label{sec:determinism}

On the \kw{-j} path, the output of the compiler does not depend on the number of threads. GIDs come from module paths and names, so they do not depend on the order of the work (Section~\ref{sec:frontend}).

CI requires the same MLIR bytes at 1 and 8 threads and with a sequential schedule. It also requires the same internal streams at 1 and 8 threads. One test came after the order of a hash map leaked into the output. Another test checks that the
output does not depend on the current directory. A grep in CI keeps locks and global atomics out of the
compiler's source. \kw{vxc --version} names the commit of a release build.

\emph{Status:} \st{tested}, with one gap. Before 22 November 2026, two files could define a macro
with the same name, and the choice changed between runs. Now a file sees only its own macros and the
macros of its imports (\#3768). The compiler reads inputs
that are not on its command line, such as environment variables (\#2250). The grep is lexical, so
it misses some forms of lazy initialization. An
interface file depends on the path of its source, because a GID hashes the module path. No test checks
that two builds of the toolchain are the same bytes.

\subsection{Tools}\label{sec:tools}

\kw{vx-format} formats a file from its tokens and leaves a file that does not lex unchanged. A configuration file can set the indent, and \kw{--check} prints a diff. The language server
\kw{vx-analyzer} publishes diagnostics as a file changes. It also answers hover, go-to-definition and
formatting requests. There is a Visual Studio Code extension and an Emacs mode. The driver \kw{vx-opt} runs the passes
of the \kw{vx} MLIR dialect, as \kw{mlir-opt} does for upstream dialects.

\emph{Status:} \kw{vx-format} and \kw{vx-opt} \st{tested}; \kw{vx-analyzer} \st{partial}. The language
server advertises find-references but does not answer it (\#2509). Its diagnostics run the parser, name resolution and the type check, but not the borrow,
placement or capacity checks. No test sends it a request (\#2510). The editor extensions have no tests in CI.

\subsection{How the project keeps its claims checkable}\label{sec:practice}

A language is maintainable only if its own documentation stays true. Vx uses five practices for this.

\paragraph{Tests that can fail.} The test suite has 2,530 Vx programs. About 1,300 must compile, and about 820 of these also run in CI. About 890 must fail with a given diagnostic, and the rest test warnings, modules and
optimizations. Of the 2,530 programs, 2,095 check their output with FileCheck. There are also about 1,400 Rust test functions.

The
project's rules ask each new test to be broken once by hand, to see that it can fail. A test comment
is a claim, and the rules ask that it be checked as one.

\paragraph{Second versions of a program.} Section~\ref{sec:principles} (P5) lists them: the CPU copy
of a kernel, the run-time twin of a \kw{comptime} block, and a differential fuzzer. The fuzzer has six
generators. It runs each generated program through both code generators and through an equivalent Rust
program, counts heap allocations, and reduces a program that disagrees. It found the bugs of six
issues, and CI runs five of its generators for 100 seeds each.

\paragraph{Formal models in CI.} A workflow builds 60 Lean files: 18,446 lines and 777 theorems,
with no \kw{sorry} and no \kw{axiom}. It runs 42 litmus tests in herd7, and it checks that Lean agrees with herd7 on the 36 tests
without branches. It also runs the TLC model checker on three TLA+ models of the runtime. A second workflow runs GenMC on the LLVM IR
that the compiler makes for four small programs. Section~\ref{sec:concurrency} says what these models
cover.

\paragraph{Documents that are checked.} A script parses every code block in the documentation and the
web site, and runs the blocks that are marked to run. The book includes 49 programs directly from the
test suite, so those cannot go stale. A script generates the index of diagnostic codes, and CI checks
that it is up to date.

\paragraph{Rules for people and AI agents.} Much of Vx is written with AI coding agents: 855 of the 3,273
commits (26\%) carry a co-author line for an AI model. The project's rules for agents and people are
the same file. A person authors every commit.

The rules ask for plain English in a controlled
style, and for a test that fails before the fix. They set a benchmark policy: absolute numbers,
the best baseline, at least three runs, and losses next to wins. Each feature must also update the
language semantics and the tutorial. The book's statements of what Vx guarantees each say where the promise stops.

\emph{Where the practice falls short:} CI has no GPU and no macOS test job. So the GPU kernels and
the Apple backends run only by hand. The job that compares the two code generators on the whole suite runs twice a week. It
began on 10 November 2026, and each of its four runs to 23 November failed. Five parser fuzz tests cannot fail, because they discard the result of the code under test
(\#3639).

The rules say that branch protection stops a push to the main branch. But the
repository does not enforce it, and four commits in November went there directly. And the README makes several claims that this report does not repeat, because the code
does not support them yet. Examples are that the frontend is ``lock-free'', that seam contracts are
``formally verified'', and that a differential suite pairs each check with CUDA. There are six such
pairs, and their CUDA half runs by hand.

\section{Secure}\label{sec:secure}

This report uses ``secure'' in a narrow sense. A language helps security when whole classes of defects cannot occur in safe code. The code that
can still have them must be marked and small. Vx aims at four classes: memory errors, data races, undefined arithmetic, and reads of memory
that a device cannot address. It also limits what each part of a program can do. In safe code, \kw{File::open} and the
socket calls of the standard library need a right. This section describes each check and this limit
on authority. Then it describes the trusted base, the security of the toolchain, and the gaps at
\kw{719726333}.

\subsection{Memory safety in safe code}\label{sec:memsafety}

The checks of Section~\ref{sec:ownership} give safe code these properties, with the exceptions of
Section~\ref{sec:secure-gaps}:
\begin{itemize}
\item no use of a value after it moves (E4001) and no write through a shared reference (E4006);
\item no exclusive reference at the same time as another reference to the same place (E4002--E4004);
\item no reference that outlives what it points to, inside a function by block regions and across
  calls by return-provenance summaries and borrow clauses (E4005, E4016);
\item no uninitialized value: every \kw{let} needs a value, and \kw{::uninit()} needs \kw{unsafe};
\item no double free: the compiler inserts each drop, and a program cannot call \kw{drop} as a
  method (E4012).
\end{itemize}

Raw pointers, foreign calls and \kw{assume} need \kw{unsafe}. The pointer clauses of
Section~\ref{sec:unsafe} let safe code call a function that takes a raw pointer, when the caller can
prove the clauses. The project translated 78 bug reports of the Rust borrow checker to Vx. They found 14
bugs, and the project fixed all of them (\#1501).

\emph{Bounds.} A constant index or a loop with constant bounds that leaves a tensor is an error at
compile time (E8003). \kw{Vec::get} always checks. Other tensor indexing is not checked at run time
unless the program is built with \kw{-fbounds-checks}, which covers CPU code only. This is the
largest gap in memory safety today (Section~\ref{sec:secure-gaps}).

\subsection{Concurrency safety}\label{sec:concurrency}

The book states the promise in one sentence: a program that compiles has no data race outside
\kw{unsafe} code. A data race is two accesses to one place at the same time, where at least one is a
write and not both are atomic. The compiler enforces the promise with the per-task borrow check of
Section~\ref{sec:tasks} and with \kw{Send} and \kw{Sync}. Table~\ref{tab:concurrency} lists how the
project checks the promise and the ones that follow from it. The Structured Tasks paper gives the
details~\cite{kumar2026tasks}.

\begin{table}[t]
\caption{What Vx promises about concurrency, and how each promise is checked. ``Model'' means a
small language or runtime that the proof is about, not the compiler.}\label{tab:concurrency}
\small
\begin{tabularx}{\linewidth}{@{}p{0.34\linewidth}Y@{}}
\toprule
Promise & Check \\
\midrule
No data race outside \kw{unsafe}. & Lean proofs for model languages with tasks, references, raw pointers, channels and \kw{Mutex}: each program that the model's checker accepts has no data
  race. \\
A program with no data race and only sequentially consistent atomics runs as if its tasks took
  turns (DRF-SC). & Lean proof for executions of any size; herd7 on one litmus test per rule; a hand-run comparison in memalloy with RC11~\cite{lahav2017rc11}, up to six events, and with
  AArch64, up to five. \\
\kw{lock}, \kw{recv} and the last drop of an \kw{Arc} see the writes that they must. & Lean proof for
  each pair; GenMC~\cite{kokologiannakis2021genmc} on the compiled LLVM IR of four programs, in every
  execution that RC11 allows. \\
A ready task does not wait forever; \kw{select} does not starve a channel. & Lean proofs about the
  scheduler queue and about \kw{select}. \\
When every task waits and nothing can wake one, the runtime reports a deadlock. & Lean proof that a deadlock stays a
  deadlock; TLC on a TLA+ model of the waits and wakes of the runtime. \\
Atomics mean the same on a GPU. & herd7 model of scopes; litmus programs in Vx run by hand on GPUs. \\
\bottomrule
\end{tabularx}
\end{table}

The proofs are about models, and the book says so. The Lean checker is not the checker in \kw{vxc}.
It uses footprints, the places that a task reads and writes, while \kw{vxc} holds whole variables. So
\kw{vxc} refuses some programs that the model accepts, and no test compares the two checkers. Only
GenMC checks code that the compiler made. The GPU litmus runs printed no weak result on an H100, but
the control tests printed none either, so those runs are weak evidence.

A run-time race detector covers the gap for code in \kw{unsafe}. The flag \kw{--sanitize=thread}
links ThreadSanitizer, and a workflow runs it every second day on each program that starts tasks.

\subsection{Arithmetic and contracts}\label{sec:arith}

Vx has no undefined integer arithmetic except a shift by the width or more. Section~\ref{sec:items}
gives the rules: wrap for \kw{+}, \kw{-} and \kw{*}, a stop for division by zero and for
\kw{usize} overflow. Because the contract checker encodes the same wrap-around, a program cannot
prove a false postcondition through an overflow (Listing~\ref{lst:contract}). A \kw{comptime} block
that overflows gives no value and is refused, so an overflow cannot fold into a wrong constant.

\subsection{Placement safety}\label{sec:placesafe}

E6003 refuses a device read of a space that the device cannot address, in safe and \kw{unsafe}
code. Two open issues are gaps in this check (\#1809, \#3638). A copy between spaces with no declared link does not compile (E6002). A pointer into one space does
not pass for a pointer into another (E6029). Capacity checks refuse a placement that the declared
machine cannot hold (E6009, E6010, E6027). These are the errors that today appear at run time as an
illegal address or an out-of-memory failure.

A small differential suite pairs six Vx programs with CUDA programs on an A100. In most
pairs, CUDA fails at run time where Vx refuses the program at compile time. The CUDA half ran once,
by hand. These checks
are only as good as the machine file, so P4 asks for a source for each figure.

\subsection{Authority}\label{sec:authority}

Memory safety stops a bug from a write to the wrong place. It does not stop a bug from an action
that the program is allowed to do. A real case shows this. The session mode of a Vx app for the
Qwen models took a file path from a request and opened it. So a request could write each file that
the process could write (qwen-vx \#105). The checks of
Sections~\ref{sec:memsafety}--\ref{sec:placesafe} do not find this bug, because each memory access
in it is correct. The app opens the file with the C function \kw{fopen}, which is trusted code. So the
rights of today do not stop this bug either. A right to one folder is designed for this case (\#3750).

The capabilities of Section~\ref{sec:capabilities} limit the code that can open a file or a socket.
The guarantee is this, apart from the limits below. Code with no \kw{unsafe}, no foreign function,
no \kw{extern impl} and no unverified \kw{mlir!} block cannot make a right. Such code calls
\kw{File::open} or opens a socket only with a right that it holds. So a request parser that gets no
right cannot call \kw{File::open} or open a socket. A right is an ordinary parameter, so a signature
shows each right that a function gets as a parameter (P2). A closure or a field of a struct can also
carry a right.

The guarantee has seven limits at \kw{719726333}:
\begin{itemize}
\item It holds only outside trusted code, and only while safe code is memory safe. A memory bug in
  \kw{unsafe} code, in a C function or in a GPU runtime can do each thing that the process can do. So
  can a bug in safe code that reaches a gap of Table~\ref{tab:gaps}.
\item Any file can declare a C function as \kw{safe} and call it with no \kw{unsafe} block. A probe
  declared the C function \kw{system} in this way, and it ran a shell command. \kw{--audit} lists
  each such declaration, and \kw{--no-trusted-code} refuses it in a package
  (Section~\ref{sec:trusted}).
\item The rights are wide. \kw{Files} opens each file that the process can open, and \kw{Network}
  opens each connection.
\item Processes, memory maps and native code have no right yet. Code reaches them through a C
  function, which is trusted code.
\item Some parts of the standard library still act with no right. Three loaders in \kw{std::llama}
  read a file at a path that they get. On a CUDA build, \kw{std::collective::join\_job} loads a
  library that an environment variable names, and it writes and reads a file that another variable
  names. Issue \#3750 tracks these parts.
\item Three gaps let safe code get a right or a handle that nobody gave it. A generic struct can be
  \kw{Copy} when a field has the type of a parameter, so a right copies (\#3861). Safe code can build
  a \kw{File} or a socket from a raw pointer, and use each descriptor that the process holds
  (\#3848). A probe wrote to standard output in this way. And a function in the file of \kw{main}
  can call \kw{main}, which makes a new \kw{Root} (\#3865).
\item A program cannot take a right back. A function can keep a lent reference to a right for as
  long as the owner holds the right. The owner is often \kw{main}, so a lent right can stay in use for
  the whole run.
\end{itemize}

\subsection{The trusted base}\label{sec:trusted}

Every Vx program trusts the compiler, its runtime and its tools. Table~\ref{tab:trusted} gives their
sizes. A program also has trusted code of its own, and \kw{--audit} lists it. Each \kw{unsafe} block,
\kw{unsafe fn}, \kw{unsafe impl}, \kw{extern impl}, foreign function and unverified \kw{mlir!} block
gets one line. Each fact that the compiler trusts without a proof also gets one line. Such a
fact is an \kw{assume} or the \kw{ensures} of a foreign function. It can also be a pointer clause of a
\kw{pub fn}, or a clause that a call in \kw{unsafe} code trusts. For a small program with one foreign function, a loop with an
\kw{unsafe} read, and two \kw{assume} statements, it prints:
\begin{lstlisting}[language={},basicstyle=\ttfamily\scriptsize]
audit.vx:4:3: trusted code: the `extern` function `c_buffer`, which runs C code
audit.vx:5:11: the `extern` function `c_buffer` ensures `valid(return, n)`, and the compiler trusts it
audit.vx:13:21: trusted code: an `unsafe` block
audit.vx:19:3: trusted code: an `unsafe` block
audit.vx:19:12: an `assume` trusts `valid(p, 4)`
audit.vx:20:3: trusted code: an `unsafe` block
audit.vx:20:12: an `assume` trusts `initialized(p, 4)`
\end{lstlisting}
A last line counts the trusted code of \kw{core} and \kw{std} in the modules that the program uses:
for this program, 10 \kw{unsafe} blocks, 4 \kw{unsafe fn} and 8 \kw{mlir!} blocks.
\kw{--audit-library} lists them too. The list does not show the \kw{unsafe} block that the compiler
makes for a \kw{main} that takes a \kw{Root}.

The flag \kw{--no-trusted-code NAME} makes each use of trusted code in the package \kw{NAME} an error
(E5003), as \kw{\#![forbid(unsafe\_code)]} does in Rust. The flag \kw{--package NAME=DIR} names the
folder of the package. The program that uses the package gives the flags. So a new version of the package cannot add trusted code without an error. Vx has no manifest
yet, so this is a flag of the compiler, not a setting of the package.

\begin{table}[t]
\caption{The trusted base at \kw{719726333}. Line counts are of tracked files; ``unsafe'' counts are of
source occurrences.}\label{tab:trusted}
\small
\begin{tabularx}{\linewidth}{@{}lY@{}}
\toprule
Part & Size and notes \\
\midrule
Compiler (Rust) & 152,249 lines in 149 files; 18 \kw{unsafe} blocks and one \kw{unsafe extern} block, mostly for the MLIR and LLVM C APIs \\
\kw{vx} MLIR dialect (C++) & 18,869 lines; the lowering pass alone is 14,941 \\
Runtimes (C, C++, Objective-C++) & 13,174 lines: CUDA, Metal, CoreML, Intel, remote worker \\
Task runtime (Rust) & 8,575 lines; 308 \kw{unsafe} blocks, none with a safety comment \\
Standard library (Vx) & 34,437 lines in 63 files; 434 \kw{unsafe} blocks, 53 \kw{unsafe fn},
  33 \kw{unsafe impl}, 8 \kw{extern impl}, about 180 \kw{mlir!} blocks, 15 \kw{extern} blocks \\
External tools & LLVM and MLIR 22.1.8 (pinned by hash), Enzyme (pinned), Z3 (any version on the
  \kw{PATH}), CUDA, Metal, CoreML, 204 Rust crates \\
\bottomrule
\end{tabularx}
\end{table}

The list covers the code of the program and its libraries. It does not cover the compiler, the
runtimes or the machine files, which the program also trusts. The design for contracts on trusted
code is still open (\#1902).

\subsection{The toolchain and deployment}\label{sec:toolchain}

The installer checks the SHA-256 sum of each release and fails closed when the sum is missing. The
sum comes from the same release, so it detects a damaged download but not a replaced release. The
releases are not signed, and there is no software bill of materials, no security policy file and no
audit of the Rust dependencies. The build pins LLVM and Enzyme to exact commits. Debug builds run the
runtime's tests under AddressSanitizer and UndefinedBehaviorSanitizer.

The task runtime supports the shadow stacks of AArch64 (GCS) and x86-64 (CET). When the operating
system turns them on, each task stack gets a shadow stack of its own. CI tests the AArch64 path
under QEMU. The compiler does not mark its output to ask for shadow stacks, so this is compatibility, not
hardening. The compiler adds no other hardening, such as stack protectors.

The remote worker listens on all interfaces over plain TCP, with no authentication. A deployment must
run it on a trusted network or through a tunnel, and the documentation does not yet say so.

\subsection{Gaps at \texorpdfstring{\kw{719726333}}{719726333}}\label{sec:secure-gaps}

This survey ran small probe programs on release builds, first at commit \kw{a06269cad} of 20
November and again at main commit \kw{719726333}. Between the two commits, the probes of three gaps
stopped. An \kw{mlir!} block that loads from an address gets E8016 (\#3790). A closure that moves a
captured value runs only once (E4001, \#3751). And only the file of a type can make it \kw{Copy}
(E3066, \#3774), so the probe of \#3743 no longer compiles. Two of these gaps remain in another form.
An \kw{mlir!} block with two operations on one line escapes E8016 (\#3864), and \#3861 copies a \kw{Vec}, so
\#3743 stays open. Table~\ref{tab:gaps} lists the places where safe code is still not safe.

\begin{table}[t]
\caption{Places where safe code is not yet safe at \kw{719726333}, from probes run on a release build.}
\label{tab:gaps}
\small
\begin{tabularx}{\linewidth}{@{}YYl@{}}
\toprule
Gap & What the probe showed & Issue \\
\midrule
A struct with \kw{Drop} holds a reference to a value that is freed first. & It reads freed memory,
  and prints a wrong value for 77 and for 7. & \#3784 \\
A \kw{match} statement on a number with no \kw{\_} arm ends a function. & The function falls off its
  end and returns 0, or a null reference that crashes. & \#3785 \\
Run-time bounds checks are off by default. & A write past the end of a tensor succeeds, then the
  program stops with SIGBUS. & \#3791 \\
\kw{@} on two \kw{[?, ?]} tensors whose inner sizes differ. & It reads past the end of a tensor,
  also with \kw{-fbounds-checks}. & \#3792 \\
A generic struct whose field has the type of a parameter, such as \kw{Twice<T>}, can be \kw{Copy}. &
  Safe code copies a \kw{Files} right. A copied \kw{Vec} reads freed memory: 6952 for 7. & \#3861 \\
Safe code builds a \kw{File} from a raw pointer. & It writes to standard output with no right,
  then the program stops. & \#3848 \\
An \kw{mlir!} block with two operations on one line, the second a load. & The check reads only the
  first operation. A safe function loads from address 16 and stops with SIGSEGV. & \#3864 \\
A function in the file of \kw{main} calls \kw{main}. & Each call makes a new \kw{Root}, with all of
  its rights. & \#3865 \\
\bottomrule
\end{tabularx}
\end{table}

Two smaller defects in the same area are open. A type can be both \kw{Drop} and \kw{Copy}, and then
its drop never runs (\#3742). And a struct invariant is a privacy rule, not a proof. So code in the
file of a type can build a value that breaks it. Section~\ref{sec:limits} lists the gaps in other
areas.

\section{Implementation}\label{sec:impl}

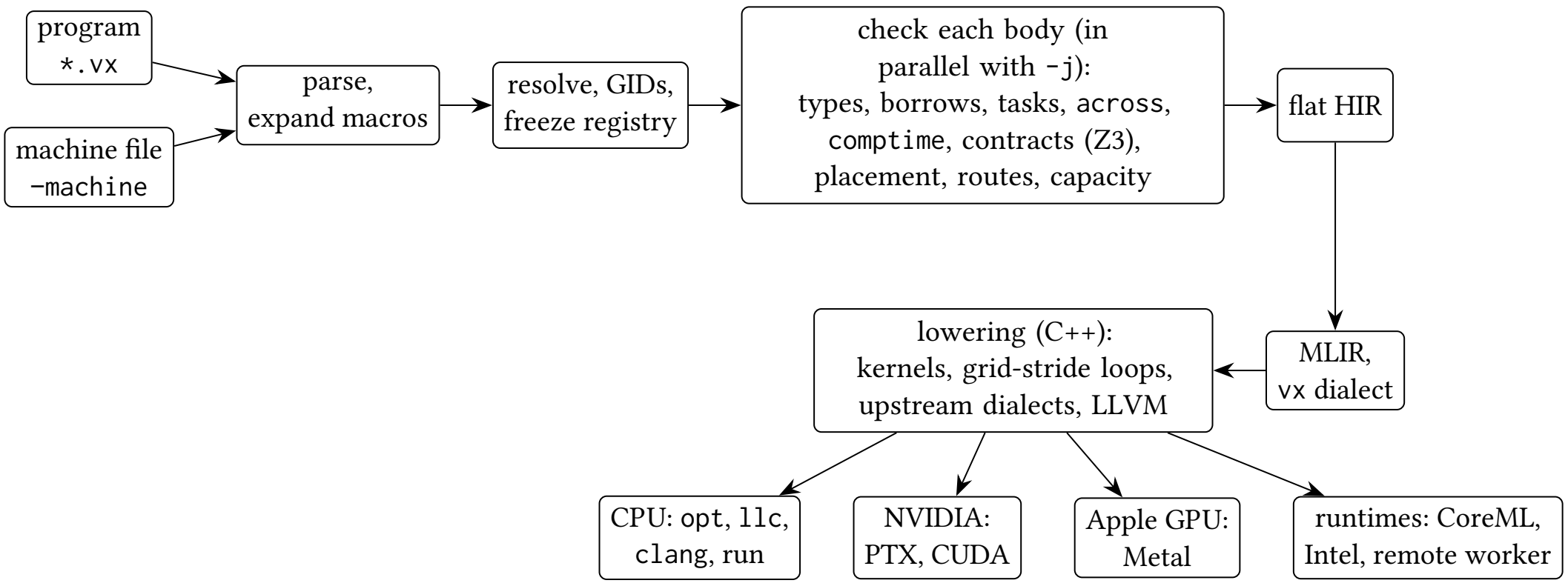
\begin{figure}[t]
\centering
\resizebox{\linewidth}{!}{%
\begin{tikzpicture}[font=\footnotesize, node distance=4mm and 5mm,
  box/.style={draw, rounded corners=2pt, align=center, inner sep=3pt, minimum height=7mm},
  arr/.style={-{Stealth[length=2mm]}}]
\node[box] (src) {program\\\kw{*.vx}};
\node[box, below=of src] (mach) {machine file\\\kw{--machine}};
\node[box, right=8mm of src, yshift=-5.5mm] (parse) {parse,\\expand macros};
\node[box, right=of parse] (freeze) {resolve, GIDs,\\freeze registry};
\node[box, right=of freeze, text width=44mm] (check) {check each body (in parallel with \kw{-j}):\\types, borrows, tasks, \kw{across},\\\kw{comptime}, contracts (Z3),\\placement, routes, capacity};
\node[box, right=of check] (hir) {flat HIR};
\node[box, below=18mm of hir] (mlir) {MLIR,\\\kw{vx} dialect};
\node[box, left=of mlir, text width=36mm] (lower) {lowering (C++):\\kernels, grid-stride loops,\\upstream dialects, LLVM};
\node[box, below=6mm of lower, xshift=-30mm] (cpu) {CPU: \kw{opt}, \kw{llc},\\\kw{clang}, run};
\node[box, right=of cpu] (ptx) {NVIDIA:\\PTX, CUDA};
\node[box, right=of ptx] (msl) {Apple GPU:\\Metal};
\node[box, right=of msl] (rt) {runtimes: CoreML,\\Intel, remote worker};
\draw[arr] (src) -- (parse); \draw[arr] (mach) -- (parse);
\draw[arr] (parse) -- (freeze); \draw[arr] (freeze) -- (check);
\draw[arr] (check) -- (hir); \draw[arr] (hir) -- (mlir); \draw[arr] (mlir) -- (lower);
\draw[arr] (lower) -- (cpu); \draw[arr] (lower) -- (ptx); \draw[arr] (lower) -- (msl); \draw[arr] (lower) -- (rt);
\end{tikzpicture}}
\caption{The path of a program through \kw{vxc}. The registry freezes before any body is checked
(P2), so the checks of different bodies do not depend on each other.}\label{fig:pipeline}
\end{figure}

The compiler, \kw{vxc}, is written in Rust. It reads a program and, when one is given, a machine file. It writes
MLIR~\cite{mlir}, LLVM IR~\cite{llvm}, object files, PTX, Metal source or SPIR-V images. This section
follows a program through it (Figure~\ref{fig:pipeline}).

\paragraph{Front end.} The lexer and parser (11,595 lines) build a syntax tree for each module. Macros
expand in their own phase, before names resolve.

The resolver gives every item a 256-bit GID
(Section~\ref{sec:frontend}). The four 64-bit words hold the module, the symbol, a generic or region
part, and flags. The first two words are FNV-1a hashes of the module path and the name, so a GID
names the same item in every run. They are not cryptographic hashes; the registry refuses two type
layouts that share a GID with different names. The registry of signatures, layouts and impls then
freezes.

\paragraph{Checker.} The checker (\kw{src/hir}, 78,301 lines) checks each function body against the
frozen registry. It types expressions, checks moves, borrows and drops, checks tasks and \kw{across}
loops, evaluates \kw{comptime} blocks, and checks placement, routes and capacity against the machine.
Its prover writes SMT-LIB queries and runs Z3 as a child process with a fixed resource limit. Then
it lowers each body to a flat stream of instructions, the flat HIR.

\paragraph{Code generation.} Two code generators make MLIR. The default one starts from the flat
HIR. The older one starts from the syntax tree. The default generator declines a construct that it
does not cover yet and hands the module to the older one. A test gate compares their output on every
pull request, and the whole suite twice a week (Section~\ref{sec:practice}).

Both emit operations of
the \kw{vx} dialect, a set of MLIR operations for tensors, transfers, device regions and lane
operations. Its lowering pass is written in C++ (18,869 lines for the dialect). It outlines each \kw{spawn on} region into a kernel, and it turns \kw{across} loops into grid-stride
loops in the \kw{gpu} dialect. It lowers the rest to the upstream dialects and to LLVM.

\paragraph{Back ends and runtimes.} For the CPU, \kw{vxc} runs \kw{mlir-translate}, \kw{opt},
\kw{llc} and \kw{clang}, then links and runs the program. Enzyme runs inside \kw{opt} when the
program uses \kw{grad}. For NVIDIA GPUs, the MLIR GPU pipeline makes PTX, and the runtime loads it through the
CUDA driver. For Apple GPUs, a printer turns the \kw{gpu} dialect into the Metal Shading Language.
The runtimes for CUDA, Metal, CoreML, Intel GPUs and remote workers are 13,174 lines of C, C++ and
Objective-C++. The task runtime, written in Rust (8,575 lines), holds the scheduler, the waits, the
task stacks and the deadlock watcher.

\paragraph{Standard library.} The library is 34,437 lines of Vx in 63 files. The module \kw{core} has options, results, iterators,
comparison, formatting and atomics. It also has a math library of 12,428 lines that the project
ported from LLVM's C library. The module \kw{std} has input and output, files, networking, time, vectors, strings,
hash maps, tensors, synchronization, collectives and \kw{std::nn}. About 180 \kw{mlir!} blocks and 15
\kw{extern} blocks connect it to MLIR operations and to C.

\paragraph{Size and history.} The repository has 3,273 commits from 14 May to 23 November 2026, by 18
author names; one author made 97\% of them. It has 3,569 tracked files: 2,651 of them are tests, and 175
are Markdown documents with 33,731 lines. Releases 0.0.1 to 0.0.5 came out between 10 September and 19
November 2026. The license is Apache 2.0 with the LLVM exception.

\section{Status and Evidence}\label{sec:status}

Table~\ref{tab:status} collects the status of each feature in this report. The status words are those
of Section~\ref{sec:intro}. The last column names the main evidence: a test directory, a hand run, a
formal model, or a companion study.

\begin{table*}[p]
\caption{The status of each feature at \kw{719726333}, and the main evidence for it.}\label{tab:status}
\footnotesize
\begin{tabularx}{\textwidth}{@{}p{0.25\textwidth}p{0.17\textwidth}Y@{}}
\toprule
Feature & Status & Evidence and limits \\
\midrule
\multicolumn{3}{@{}l}{\emph{Core language}} \\
Types, numbers, structs, enums, tensors & \st{tested} & frontend and backend tests; 128-bit print fails \\
Pattern matching & \st{partial} & no guards, or-patterns or nested patterns; \#3785 \\
Generics, traits & \st{tested} & impls in the file of the type (E3066); \kw{Copy} gaps (\#3861, \#3742); impl order (\#3787) \\
Closures & \st{tested} & a closure that moves a capture runs once (E4001) \\
Moves, borrows, drops & \st{tested} & 130 failing borrow tests; \#1501; \#3784 \\
Borrow clauses & \st{tested} & E4016 tests; \kw{--print-borrow-clauses} \\
Linear values & \st{designed} & the code is affine \\
\kw{comptime}, macros & \st{tested} & 13 run-time twins in CI; Must-Fold generator~\cite{kumar2026mustfold} \\
Contracts, loop invariants & \st{tested} & 137 programs use them; wrap-aware encoding \\
Struct invariants, \kw{Verified} & \st{partial} & privacy rule; not proved \\
\kw{unsafe}, pointer clauses, FFI & \st{tested} & \kw{safe} foreign functions need no \kw{unsafe} at a call \\
\kw{mlir!} & \st{tested} & outside \kw{unsafe}, only verified blocks (E8016); two operations on one line escape the check (\#3864) \\
Capabilities: \kw{Files}, \kw{Network} & \st{tested} & forge tests; \kw{Twice<T>} copies one (\#3861); \#3848; \kw{std::llama} \\
Right to one folder & \st{designed} & \#3750; no design yet for processes, memory maps or native code \\
\midrule
\multicolumn{3}{@{}l}{\emph{Scalable}} \\
Machine files & \st{tested} & 16 files; 257 test files use \kw{--machine}; provenance unchecked \\
Placement, visibility, routes & \st{tested} & E6003 in 21 tests; routes against Bellman--Ford \\
Capacity, peaks, cross-call & \st{tested} & 24/24 launches on two GPUs~\cite{kumar2026admission}; \#3632 \\
Seam checks & \st{tested}, off by default & three tests; decides only ``relaxed or not'' (\#662) \\
User transfer bodies & \st{partial} & only edges into multiprocessor memory \\
\kw{spawn on} & \st{tested} & per backend; device 0 only (E6042) \\
Tasks, channels, \kw{Mutex}, \kw{Arc}, atomics & \st{tested} & x86-64 in CI; runs under three runtimes \\
\kw{select} & \st{tested} & sources of different types: channels, sockets, \kw{Semaphore}, \kw{Interval}, \kw{Deadline} \\
Deadlock watcher & \st{partial} & reports only when every thread waits; names the sources of a \kw{select} \\
\kw{across} loops, race and barrier rules & \st{tested} & static checks; 38-kernel corpus~\cite{kumar2026erasable} \\
Kernels on NVIDIA GPUs & \st{tested by hand} & A100, A6000, H100, H200 runs; PTX text in CI \\
Kernels on Apple GPUs & \st{tested by hand} & macOS only \\
Apple neural engine & \st{partial} & four prebuilt models; other regions on the CPU; \#3793 \\
Intel GPU & \st{partial} & \kw{across} loops, lane routines and atomics in SPIR-V; one card run; no CI \\
AMD GPU, Cortex-M7 & \st{designed} & machine files only \\
Tensors, \kw{@} & \st{tested} & inner dimension of \kw{[?, ?]} unchecked (\#3792) \\
Automatic differentiation & \st{tested} & scalars only \\
\kw{std::nn} & \st{tested by hand} & Apple GPU; CPU tests in CI \\
Collectives & \st{partial} & one rank \\
Parallel frontend, determinism & \st{tested} & byte-identical output gates; 3.9--5.9$\times$ on 48 cores~\cite{kumar2026frontend}; opt-in \\
Remote workers & \st{tested}, loopback & one two-machine run; no authentication \\
NUMA binding & \st{tested by hand} & one \kw{c5.metal} run \\
\midrule
\multicolumn{3}{@{}l}{\emph{Maintainable}} \\
Modules, interfaces & \st{tested} & interface depends on the source path \\
\kw{vx-api} & \st{tested} & 17 tests; blind to contracts and visibility \\
Visibility levels & \st{designed} & only \kw{pub fn} export today \\
Package manager & \st{designed} & \#1275 \\
Diagnostics, JSON, error pages & \st{tested} & 196 codes, 22 never emitted \\
\kw{vx-format}, \kw{vx-opt} & \st{tested} & \\
\kw{vx-analyzer} & \st{partial} & no borrow or placement diagnostics; \#2509, \#2510 \\
\midrule
\multicolumn{3}{@{}l}{\emph{Secure}} \\
Race freedom, DRF-SC & proved for a model & Lean (777 theorems in all), herd7, memalloy; not connected to \kw{vxc} \\
Synchronization orders & proved for a model; GenMC & GenMC on compiled IR of four programs \\
Run-time bounds checks & opt-in & \kw{-fbounds-checks}, CPU only (\#3791) \\
\kw{--audit}, \kw{--no-trusted-code} & \st{tested} & no manifest; a flag of the compiler \\
Thread sanitizer & \st{tested} & every second day \\
Shadow stacks & \st{partial} & compatibility only; AArch64 under QEMU \\
Release integrity & \st{partial} & SHA-256 sums; no signing \\
\bottomrule
\end{tabularx}
\end{table*}

\paragraph{The companion studies.} Five studies test the claims of this report on real hardware or at
scale, and each one states its own limits.
\begin{itemize}
\item \emph{Capacity on real GPUs.} The compiler predicted, before any rental, which of 12 LLM serving
  configurations fit on an H100 and an A100. All 24 launches agreed with the predictions. The engine
  needed 1.1 to 1.9~GiB more than the prediction, and a fixed headroom of 7\% covered every admitted
  launch~\cite{kumar2026admission}.
\item \emph{Machine files against hardware.} A study froze the predictions of the machine files, then rented
  the hardware and measured it. It reported where the declarations were wrong, such as the missing
  startup latency of a small copy~\cite{kumar2026machinemodel}.
\item \emph{The parallel frontend.} The frontend gave the same output at every thread count. An August build
  reached a speedup of 5.9 on 48 cores, to MLIR text; a September run gave 3.9 to
  4.5~\cite{kumar2026frontend}.
\item \emph{Compile-time evaluation.} A generator made 57,950 programs with \kw{comptime} blocks, and none folded to a wrong value~\cite{kumar2026mustfold}.
\item \emph{GPU loop nests.} The study rewrote 38 kernels from three benchmark suites as
  \kw{across} loops. The checker accepted 27 and refused 11, and three of the refused kernels had
  real races~\cite{kumar2026erasable}.
\end{itemize}

\section{Related Work}\label{sec:related}

Vx combines ideas from four lines of work. The first two are systems languages with ownership and languages for parallel and heterogeneous
machines. The others are verification inside a language, and tools that keep the interfaces and
claims of a language checkable. This section places Vx in each line.

\paragraph{Systems languages.} Rust~\cite{matsakis2014rust} brought ownership and borrowing to a
systems language, and RustBelt gave its type system a formal basis~\cite{jung2018rustbelt}. Vx takes its borrow rules from Rust, and it differs in three places. First, it has no lifetime
syntax, and it reads from the callee where a returned reference may point
(Section~\ref{sec:ownership}). Second, it has no \kw{dyn}, no \kw{impl Trait} result and no
specialization, so that a signature is the whole contract. Third, its integer arithmetic is defined in
every build, except a shift by the width of the type or more.

Zig's \kw{comptime}~\cite{zigdocs} and C++ \kw{constexpr}~\cite{dosreis2010constexpr} run code in
the compiler. Like Zig, Vx refuses a block that it cannot fold. Vx also
lets a block name a run-time value when the result does not depend on it. The closest prior work is hybrid partial evaluation~\cite{shali2011hybrid}
and staging~\cite{taha1997multistage}.

Mojo~\cite{mojo} aims at systems code for new hardware, and Carbon~\cite{carbon} at a successor to
C++. Mojo puts an address space on pointer types, but neither language puts capacities or machine
declarations in the program. Tools such as Miri~\cite{jung2026miri} find undefined
behavior in \kw{unsafe} Rust at run time. Vx has no such interpreter yet, and its \kw{--audit}
flag is a static list instead.

\paragraph{Parallel and heterogeneous languages.} CUDA~\cite{cudaguide}, OpenCL~\cite{opencl} and
SYCL~\cite{sycl2020} put address spaces on pointers. But a pointer into device memory and one into host memory still mix
at run time, and capacities are not part of the program. Chapel's locales~\cite{chapel,
chapel2021localemodels} and X10's places~\cite{x10} describe where data lives in a cluster. Legion and
Regent~\cite{legion,regent} map logical regions to memories at run time with a mapper, and
Sequoia~\cite{sequoia} made the memory hierarchy part of the program. Kokkos~\cite{kokkos} gives
portable parallel patterns in C++. Vx shares the goal of these systems, but it checks placement,
visibility and capacity at compile time against a machine that a source file declares.

Among kernel languages, Halide~\cite{ragankelley13} separates the algorithm from its schedule,
Triton~\cite{tillet19} programs tiles, and TVM~\cite{chen2018tvm} compiles tensor programs.
Exo~\cite{exo} and Exo-GPU~\cite{akeley2026exogpu} let a programmer rewrite a schedule under checks. Exo-GPU checks the
equivalence of a schedule by interpreting it at problem sizes that the programmer chooses, and the
programmer writes the synchronization. Descend~\cite{descend} is the closest kernel language. Its types track memory spaces
and the levels of the thread hierarchy, and its borrow checker rules out races in a kernel. Vx's \kw{across} loops keep the sequential meaning of the loop nest instead, and the compiler
inserts the barriers.

Futhark~\cite{henriksen2017futhark} compiles a functional language with nested
parallelism to GPUs. GPUVerify~\cite{betts2012gpuverify} and Faial~\cite{cogumbreiro2021faial} check kernels for races
after a programmer writes them. Vx builds a decidable part of the same check into the language and
refuses the rest.

Deterministic Parallel Java~\cite{bocchino2009type} uses effects to show that parallel tasks do not
interfere. Habanero-Java~\cite{cave2011habanero} and Cilk~\cite{frigo1998implementation} give tasks
that cannot outlive their scope, as Vx \kw{tasks} blocks do. ML5~\cite{ml5} and the modal
lambda calculus of Murphy et al.~\cite{lambda5} type where code runs in a distributed program.

\paragraph{Verification in a language.} Dafny~\cite{leino2010dafny}, F*~\cite{swamy2016fstar} and
Liquid Haskell~\cite{vazou2014refinement} check contracts or refinement types with an SMT solver. For
Rust, Prusti~\cite{astrauskas2019prusti}, Creusot~\cite{denis2022creusot}, Verus~\cite{lattuada2023verus}
and Kani~\cite{vanhattum2022kani} add specifications or bounded model checking. Vx's contracts are
weaker than these systems: they cover integer facts and pointer clauses, not general functional
correctness. But they are part of the compiler that every program uses, and their encoding models the
same wrap-around as the generated code. CompCert~\cite{leroy2009compcert} proves a compiler correct;
Vx proves properties of models
(Section~\ref{sec:concurrency}), and only GenMC checks compiler output against one of them. That is
a much smaller claim. The memory-model work that Vx builds on is RC11~\cite{lahav2017rc11},
herd7~\cite{alglave2014herding}, memalloy~\cite{wickerson2017memalloy} and
GenMC~\cite{kokologiannakis2021genmc}, and its proofs are in Lean~4~\cite{demoura2021lean4}.

\paragraph{Capabilities.} Dennis and Van Horn named capabilities~\cite{dennis1966capabilities}.
Miller's object-capability model asks that authority moves only through references that a
program holds~\cite{miller2006robust}. Joe-E~\cite{mettler2010joee} and Wyvern~\cite{melicher2017capability}
limit a language, so that a module gets authority only from the code that uses it. Pony gives
\kw{main} a value of ambient authority, from which a program makes the right to use files~\cite{ponycaps}.
Austral gives a root capability to the function that starts the program~\cite{austral}. Capsicum~\cite{watson2010capsicum},
WASI~\cite{wasi} and the Rust crate cap-std~\cite{capstd} replace paths with handles to folders
that the process already holds. Vx takes the root capability of Pony and Austral. Like Austral, it makes each right a type that only
one module can build. Austral's capabilities are linear, so a program must use each one. Vx's rights
are affine, so a program can drop one with no error. Joe-E and Wyvern remove each path to the operating
system that does not go through a capability. Vx does not do this yet. Trusted code can still call
the operating system, and \kw{--audit} lists that code (Section~\ref{sec:authority}).

\paragraph{Interfaces, versions and testing.} Elm's package tool computes the version change of a release from its
types~\cite{elmpackage}, and \kw{cargo-semver-checks} does so for Rust~\cite{cargosemverchecks}.
The tool \kw{vx-api} does the same from a frozen interface. The differential
testing of the two code generators follows McKeeman~\cite{mckeeman1998differential} and
Csmith~\cite{yang2011csmith}. Automatic differentiation uses Enzyme~\cite{moses2020enzyme}, and the
contracts use Z3~\cite{demoura2008z3}.

\section{Limitations and Future Work}\label{sec:limits}

This report describes one commit of a young language, and most of its limits are already stated next
to the features. This section collects the ones that matter most.

\paragraph{Safety.} Table~\ref{tab:gaps} lists the places where safe code is not yet safe. The most important design decision that remains is the default for run-time bounds checks
(\#3791). Other open items in
this area are the \kw{Copy} gaps (\#3861, \#3742), proofs of struct invariants, and contracts on
trusted code (\#1902).

\paragraph{Authority.} Files and sockets need a right, but the rights are wide, and a program cannot
take one back. Processes, memory maps and native code have no right yet, and some loaders of the standard library
read files with no right. A foreign function that is
declared \kw{safe} reaches the operating system with no right and no \kw{unsafe} at its calls.

\paragraph{Proofs and the compiler.} The Lean proofs, herd7 models and TLA+ models are about models,
not about \kw{vxc}. The task checker of the compiler holds whole variables, while the model uses
footprints, and no test compares them. There is no formal model of placement, transfers, capacity
or seams (\#1209), and no mechanized proof of the race rule of \kw{across} loops. The seam check decides
only whether a hop is relaxed (\#662).

\paragraph{Hardware coverage.} CI runs on x86-64 Linux CPUs. The GPU backends, the Apple backends and the NUMA binding run only by hand. A test runs the SPIR-V images of the Intel GPU tests on pocl, an OpenCL for CPUs, but CI has no pocl yet (\#3731). Most hand runs are recorded only in commit messages and issues, and no file of test
results holds them. The neural engine runs four prebuilt models. AMD and microcontroller targets have machine
files but no code generator. The cost model has no latency per transfer and no contention.

\paragraph{Language and tools.} Pattern matching lacks guards, or-patterns and nested patterns.
Visibility has one level, there is no package manager, and \kw{vx-api} does not compare contracts. The
language server does not run the borrow or placement checks. The parallel frontend is opt-in, and some
flags do not work with it. A machine file cannot import another.

\paragraph{Future work.} The next steps follow from these limits:
\begin{itemize}
\item make bounds checks safe by default;
\item connect the task checker to its Lean model with a differential test, then move it to footprints;
\item give placement and capacity a formal model;
\item run the GPU and macOS suites in CI on real hardware;
\item add visibility levels and a package manager;
\item add a right to one folder, and rights for processes and native code;
\item make each call of a \kw{safe} foreign function easy to find in a review.
\end{itemize}

\section{Conclusion}\label{sec:conclusion}

Vx makes three kinds of facts part of the program text. The first is where data lives and which
machine runs the code. The second is what a function promises to its callers. The third is what
the compiler trusts without a proof.

Because these facts are in the text, the compiler can check them before the program runs.
It can also check function bodies in parallel, and say exactly what a change broke. That is the common root of the three qualities of the
title. The language is scalable because one text runs across machines, cores and GPU threads. It is
maintainable because a signature is the contract of a function, with one stated exception, and
diagnostics name what failed. It
is secure because safe code has ownership, race and placement checks. And in safe code, \kw{File::open} and
the socket calls of the standard library need a right. A flag lists the trusted code of the program.

The report also shows where each quality is not yet met. We think that a language should be
described with its gaps, so that a reader can test the claims against the code.

\bibliographystyle{ACM-Reference-Format}
\bibliography{refs}


\begin{thebibliography}{72}


\ifx \showCODEN    \undefined \def \showCODEN     #1{\unskip}     \fi
\ifx \showISBNx    \undefined \def \showISBNx     #1{\unskip}     \fi
\ifx \showISBNxiii \undefined \def \showISBNxiii  #1{\unskip}     \fi
\ifx \showISSN     \undefined \def \showISSN      #1{\unskip}     \fi
\ifx \showLCCN     \undefined \def \showLCCN      #1{\unskip}     \fi
\ifx \shownote     \undefined \def \shownote      #1{#1}          \fi
\ifx \showarticletitle \undefined \def \showarticletitle #1{#1}   \fi
\ifx \showURL      \undefined \def \showURL       {\relax}        \fi
\providecommand\bibfield[2]{#2}
\providecommand\bibinfo[2]{#2}
\providecommand\natexlab[1]{#1}
\providecommand\showeprint[2][]{arXiv:#2}

\bibitem[Akeley et~al\mbox{.}(2026)]%
        {akeley2026exogpu}
\bibfield{author}{\bibinfo{person}{David~Zhao Akeley}, \bibinfo{person}{Yuka
  Ikarashi}, {and} \bibinfo{person}{Jonathan Ragan-Kelley}.}
  \bibinfo{year}{2026}\natexlab{}.
\newblock \bibinfo{title}{{Exo-GPU}: Safe, Imperative, User-schedulable
  Programming for Tensor Cores}.
\newblock
\showeprint[arxiv]{2609.16389}~[cs.PL]
\urldef\tempurl%
\url{https://arxiv.org/abs/2609.16389}
\showURL{%
\tempurl}
\newblock
\shownote{Posted 2026-09-14; arXiv comment: ``no conference publication''}.


\bibitem[Alglave et~al\mbox{.}(2014)]%
        {alglave2014herding}
\bibfield{author}{\bibinfo{person}{Jade Alglave}, \bibinfo{person}{Luc
  Maranget}, {and} \bibinfo{person}{Michael Tautschnig}.}
  \bibinfo{year}{2014}\natexlab{}.
\newblock \showarticletitle{Herding Cats: Modelling, Simulation, Testing, and
  Data Mining for Weak Memory}.
\newblock \bibinfo{journal}{\emph{ACM Transactions on Programming Languages and
  Systems}} \bibinfo{volume}{36}, \bibinfo{number}{2} (\bibinfo{year}{2014}),
  \bibinfo{pages}{1--74}.
\newblock
\href{https://doi.org/10.1145/2627752}{doi:\nolinkurl{10.1145/2627752}}


\bibitem[Astrauskas et~al\mbox{.}(2019)]%
        {astrauskas2019prusti}
\bibfield{author}{\bibinfo{person}{Vytautas Astrauskas}, \bibinfo{person}{Peter
  M{\"u}ller}, \bibinfo{person}{Federico Poli}, {and}
  \bibinfo{person}{Alexander~J. Summers}.} \bibinfo{year}{2019}\natexlab{}.
\newblock \showarticletitle{Leveraging {Rust} Types for Modular Specification
  and Verification}.
\newblock \bibinfo{journal}{\emph{Proceedings of the ACM on Programming
  Languages}} \bibinfo{volume}{3}, \bibinfo{number}{OOPSLA}
  (\bibinfo{year}{2019}), \bibinfo{pages}{1--30}.
\newblock
\href{https://doi.org/10.1145/3360573}{doi:\nolinkurl{10.1145/3360573}}


\bibitem[Bauer et~al\mbox{.}(2012)]%
        {legion}
\bibfield{author}{\bibinfo{person}{Michael Bauer}, \bibinfo{person}{Sean
  Treichler}, \bibinfo{person}{Elliott Slaughter}, {and} \bibinfo{person}{Alex
  Aiken}.} \bibinfo{year}{2012}\natexlab{}.
\newblock \showarticletitle{Legion: Expressing Locality and Independence with
  Logical Regions}. In \bibinfo{booktitle}{\emph{SC}}.
\newblock
\href{https://doi.org/10.1109/SC.2012.71}{doi:\nolinkurl{10.1109/SC.2012.71}}


\bibitem[Betts et~al\mbox{.}(2012)]%
        {betts2012gpuverify}
\bibfield{author}{\bibinfo{person}{Adam Betts}, \bibinfo{person}{Nathan Chong},
  \bibinfo{person}{Alastair Donaldson}, \bibinfo{person}{Shaz Qadeer}, {and}
  \bibinfo{person}{Paul Thomson}.} \bibinfo{year}{2012}\natexlab{}.
\newblock \showarticletitle{{GPUVerify}: A Verifier for {GPU} Kernels}. In
  \bibinfo{booktitle}{\emph{Proceedings of the ACM International Conference on
  Object Oriented Programming Systems Languages and Applications (OOPSLA
  '12)}}. \bibinfo{publisher}{ACM}, \bibinfo{pages}{113--132}.
\newblock
\href{https://doi.org/10.1145/2384616.2384625}{doi:\nolinkurl{10.1145/2384616.2384625}}


\bibitem[Bocchino et~al\mbox{.}(2009)]%
        {bocchino2009type}
\bibfield{author}{\bibinfo{person}{Robert~L. Bocchino},
  \bibinfo{person}{Vikram~S. Adve}, \bibinfo{person}{Danny Dig},
  \bibinfo{person}{Sarita~V. Adve}, \bibinfo{person}{Stephen Heumann},
  \bibinfo{person}{Rakesh Komuravelli}, \bibinfo{person}{Jeffrey Overbey},
  \bibinfo{person}{Patrick Simmons}, \bibinfo{person}{Hyojin Sung}, {and}
  \bibinfo{person}{Mohsen Vakilian}.} \bibinfo{year}{2009}\natexlab{}.
\newblock \showarticletitle{A Type and Effect System for {Deterministic
  Parallel Java}}. In \bibinfo{booktitle}{\emph{Proceedings of the 24th ACM
  SIGPLAN Conference on Object Oriented Programming Systems Languages and
  Applications (OOPSLA)}}. \bibinfo{publisher}{ACM}, \bibinfo{pages}{97--116}.
\newblock
\href{https://doi.org/10.1145/1640089.1640097}{doi:\nolinkurl{10.1145/1640089.1640097}}


\bibitem[Borretti(2026)]%
        {austral}
\bibfield{author}{\bibinfo{person}{Fernando Borretti}.}
  \bibinfo{year}{2026}\natexlab{}.
\newblock \bibinfo{title}{The {Austral} Language Specification}.
\newblock
  \bibinfo{howpublished}{\url{https://austral-lang.org/spec/spec.html}}.
\newblock
\newblock
\shownote{Accessed 2026-11-24}.


\bibitem[{Bytecode Alliance}(2026)]%
        {capstd}
\bibfield{author}{\bibinfo{person}{{Bytecode Alliance}}.}
  \bibinfo{year}{2026}\natexlab{}.
\newblock \bibinfo{title}{cap-std: A Capability-Based Version of the {Rust}
  Standard Library}.
\newblock
  \bibinfo{howpublished}{\url{https://github.com/bytecodealliance/cap-std}}.
\newblock
\newblock
\shownote{Accessed 2026-11-24}.


\bibitem[Cav{\'e} et~al\mbox{.}(2011)]%
        {cave2011habanero}
\bibfield{author}{\bibinfo{person}{Vincent Cav{\'e}}, \bibinfo{person}{Jisheng
  Zhao}, \bibinfo{person}{Jun Shirako}, {and} \bibinfo{person}{Vivek Sarkar}.}
  \bibinfo{year}{2011}\natexlab{}.
\newblock \showarticletitle{{Habanero-Java}: The New Adventures of Old {X10}}.
  In \bibinfo{booktitle}{\emph{Proceedings of the 9th International Conference
  on Principles and Practice of Programming in Java (PPPJ)}}.
  \bibinfo{publisher}{ACM}, \bibinfo{pages}{51--61}.
\newblock
\href{https://doi.org/10.1145/2093157.2093165}{doi:\nolinkurl{10.1145/2093157.2093165}}


\bibitem[Chamberlain et~al\mbox{.}(2007)]%
        {chapel}
\bibfield{author}{\bibinfo{person}{Bradford~L. Chamberlain},
  \bibinfo{person}{David Callahan}, {and} \bibinfo{person}{Hans~P. Zima}.}
  \bibinfo{year}{2007}\natexlab{}.
\newblock \showarticletitle{Parallel Programmability and the Chapel Language}.
\newblock \bibinfo{journal}{\emph{International Journal of High Performance
  Computing Applications}} \bibinfo{volume}{21}, \bibinfo{number}{3}
  (\bibinfo{year}{2007}).
\newblock
\href{https://doi.org/10.1177/1094342007078442}{doi:\nolinkurl{10.1177/1094342007078442}}


\bibitem[{Chapel project}(2021)]%
        {chapel2021localemodels}
\bibfield{author}{\bibinfo{person}{{Chapel project}}.}
  \bibinfo{year}{2021}\natexlab{}.
\newblock \bibinfo{title}{Chapel 1.25 Technical Note: Locale Models}.
\newblock
\urldef\tempurl%
\url{https://chapel-lang.org/docs/1.25/technotes/localeModels.html}
\showURL{%
\tempurl}


\bibitem[Charles et~al\mbox{.}(2005)]%
        {x10}
\bibfield{author}{\bibinfo{person}{Philippe Charles},
  \bibinfo{person}{Christian Grothoff}, \bibinfo{person}{Vijay Saraswat},
  \bibinfo{person}{Christopher Donawa}, \bibinfo{person}{Allan Kielstra},
  \bibinfo{person}{Kemal Ebcio{\u{g}}lu}, \bibinfo{person}{Christoph von
  Praun}, {and} \bibinfo{person}{Vivek Sarkar}.}
  \bibinfo{year}{2005}\natexlab{}.
\newblock \showarticletitle{X10: An Object-Oriented Approach to Non-Uniform
  Cluster Computing}. In \bibinfo{booktitle}{\emph{OOPSLA}}.
\newblock
\href{https://doi.org/10.1145/1094811.1094852}{doi:\nolinkurl{10.1145/1094811.1094852}}


\bibitem[Chen et~al\mbox{.}(2018)]%
        {chen2018tvm}
\bibfield{author}{\bibinfo{person}{Tianqi Chen}, \bibinfo{person}{Thierry
  Moreau}, \bibinfo{person}{Ziheng Jiang}, \bibinfo{person}{Lianmin Zheng},
  \bibinfo{person}{Eddie Yan}, \bibinfo{person}{Haichen Shen},
  \bibinfo{person}{Meghan Cowan}, \bibinfo{person}{Leyuan Wang},
  \bibinfo{person}{Yuwei Hu}, \bibinfo{person}{Luis Ceze},
  \bibinfo{person}{Carlos Guestrin}, {and} \bibinfo{person}{Arvind
  Krishnamurthy}.} \bibinfo{year}{2018}\natexlab{}.
\newblock \showarticletitle{{TVM}: An Automated End-to-End Optimizing Compiler
  for Deep Learning}. In \bibinfo{booktitle}{\emph{13th USENIX Symposium on
  Operating Systems Design and Implementation (OSDI 18)}}.
  \bibinfo{publisher}{USENIX Association}, \bibinfo{pages}{578--594}.
\newblock
\showISBNx{978-1-939133-08-3}
\urldef\tempurl%
\url{https://www.usenix.org/conference/osdi18/presentation/chen}
\showURL{%
\tempurl}


\bibitem[Cogumbreiro et~al\mbox{.}(2021)]%
        {cogumbreiro2021faial}
\bibfield{author}{\bibinfo{person}{Tiago Cogumbreiro}, \bibinfo{person}{Julien
  Lange}, \bibinfo{person}{Dennis Liew Zhen~Rong}, {and}
  \bibinfo{person}{Hannah Zicarelli}.} \bibinfo{year}{2021}\natexlab{}.
\newblock \showarticletitle{Checking Data-Race Freedom of {GPU} Kernels,
  Compositionally}.
\newblock In \bibinfo{booktitle}{\emph{Computer Aided Verification (CAV
  2021)}}, \bibfield{editor}{\bibinfo{person}{Alexandra Silva} {and}
  \bibinfo{person}{K.~Rustan~M. Leino}} (Eds.). \bibinfo{publisher}{Springer},
  \bibinfo{pages}{403--426}.
\newblock
\href{https://doi.org/10.1007/978-3-030-81685-8_19}{doi:\nolinkurl{10.1007/978-3-030-81685-8_19}}


\bibitem[de~Moura and Bj{\o}rner(2008)]%
        {demoura2008z3}
\bibfield{author}{\bibinfo{person}{Leonardo de Moura} {and}
  \bibinfo{person}{Nikolaj Bj{\o}rner}.} \bibinfo{year}{2008}\natexlab{}.
\newblock \showarticletitle{{Z3}: An Efficient {SMT} Solver}. In
  \bibinfo{booktitle}{\emph{Tools and Algorithms for the Construction and
  Analysis of Systems (TACAS)}} \emph{(\bibinfo{series}{LNCS},
  Vol.~\bibinfo{volume}{4963})}. \bibinfo{publisher}{Springer},
  \bibinfo{address}{Berlin, Heidelberg}, \bibinfo{pages}{337--340}.
\newblock
\href{https://doi.org/10.1007/978-3-540-78800-3_24}{doi:\nolinkurl{10.1007/978-3-540-78800-3_24}}


\bibitem[de~Moura and Ullrich(2021)]%
        {demoura2021lean4}
\bibfield{author}{\bibinfo{person}{Leonardo de Moura} {and}
  \bibinfo{person}{Sebastian Ullrich}.} \bibinfo{year}{2021}\natexlab{}.
\newblock \showarticletitle{The {Lean} 4 Theorem Prover and Programming
  Language}. In \bibinfo{booktitle}{\emph{Automated Deduction -- CADE 28}}
  \emph{(\bibinfo{series}{Lecture Notes in Computer Science},
  Vol.~\bibinfo{volume}{12699})}. \bibinfo{publisher}{Springer},
  \bibinfo{pages}{625--635}.
\newblock
\href{https://doi.org/10.1007/978-3-030-79876-5_37}{doi:\nolinkurl{10.1007/978-3-030-79876-5_37}}


\bibitem[Denis et~al\mbox{.}(2022)]%
        {denis2022creusot}
\bibfield{author}{\bibinfo{person}{Xavier Denis},
  \bibinfo{person}{Jacques-Henri Jourdan}, {and} \bibinfo{person}{Claude
  March{\'e}}.} \bibinfo{year}{2022}\natexlab{}.
\newblock \showarticletitle{Creusot: A Foundry for the Deductive Verification
  of {Rust} Programs}. In \bibinfo{booktitle}{\emph{Formal Methods and Software
  Engineering (ICFEM)}}. \bibinfo{publisher}{Springer},
  \bibinfo{pages}{90--105}.
\newblock
\href{https://doi.org/10.1007/978-3-031-17244-1_6}{doi:\nolinkurl{10.1007/978-3-031-17244-1_6}}


\bibitem[Dennis and Horn(1966)]%
        {dennis1966capabilities}
\bibfield{author}{\bibinfo{person}{Jack~B. Dennis} {and} \bibinfo{person}{Earl
  C.~Van Horn}.} \bibinfo{year}{1966}\natexlab{}.
\newblock \showarticletitle{Programming Semantics for Multiprogrammed
  Computations}.
\newblock \bibinfo{journal}{\emph{Commun. ACM}} \bibinfo{volume}{9},
  \bibinfo{number}{3} (\bibinfo{year}{1966}), \bibinfo{pages}{143--155}.
\newblock
\href{https://doi.org/10.1145/365230.365252}{doi:\nolinkurl{10.1145/365230.365252}}


\bibitem[{Dos Reis} and Stroustrup(2010)]%
        {dosreis2010constexpr}
\bibfield{author}{\bibinfo{person}{Gabriel {Dos Reis}} {and}
  \bibinfo{person}{Bjarne Stroustrup}.} \bibinfo{year}{2010}\natexlab{}.
\newblock \showarticletitle{General Constant Expressions for System Programming
  Languages}. In \bibinfo{booktitle}{\emph{Proceedings of the 2010 ACM
  Symposium on Applied Computing (SAC)}}. \bibinfo{publisher}{ACM},
  \bibinfo{address}{New York, NY, USA}, \bibinfo{pages}{2131--2136}.
\newblock
\href{https://doi.org/10.1145/1774088.1774537}{doi:\nolinkurl{10.1145/1774088.1774537}}


\bibitem[Edwards et~al\mbox{.}(2014)]%
        {kokkos}
\bibfield{author}{\bibinfo{person}{H.~Carter Edwards},
  \bibinfo{person}{Christian~R. Trott}, {and} \bibinfo{person}{Daniel
  Sunderland}.} \bibinfo{year}{2014}\natexlab{}.
\newblock \showarticletitle{Kokkos: Enabling Manycore Performance Portability
  Through Polymorphic Memory Access Patterns}.
\newblock \bibinfo{journal}{\emph{J. Parallel and Distrib. Comput.}}
  \bibinfo{volume}{74}, \bibinfo{number}{12} (\bibinfo{year}{2014}).
\newblock
\href{https://doi.org/10.1016/j.jpdc.2014.07.003}{doi:\nolinkurl{10.1016/j.jpdc.2014.07.003}}


\bibitem[{Elm}(2026)]%
        {elmpackage}
\bibfield{author}{\bibinfo{person}{{Elm}}.} \bibinfo{year}{2026}\natexlab{}.
\newblock \bibinfo{title}{elm-package: Command Line Tool to Share {Elm}
  Libraries}.
\newblock
  \bibinfo{howpublished}{\url{https://github.com/elm-lang/elm-package}}.
\newblock
\newblock
\shownote{README; superseded by the \texttt{elm bump}/\texttt{elm diff}
  commands of Elm 0.19. Accessed 2026-10-16}.


\bibitem[Fatahalian et~al\mbox{.}(2006)]%
        {sequoia}
\bibfield{author}{\bibinfo{person}{Kayvon Fatahalian},
  \bibinfo{person}{Daniel~Reiter Horn}, \bibinfo{person}{Timothy~J. Knight},
  \bibinfo{person}{Larkhoon Leem}, \bibinfo{person}{Mike Houston},
  \bibinfo{person}{Ji~Young Park}, \bibinfo{person}{Mattan Erez},
  \bibinfo{person}{Manman Ren}, \bibinfo{person}{Alex Aiken},
  \bibinfo{person}{William~J. Dally}, {and} \bibinfo{person}{Pat Hanrahan}.}
  \bibinfo{year}{2006}\natexlab{}.
\newblock \showarticletitle{Sequoia: Programming the Memory Hierarchy}. In
  \bibinfo{booktitle}{\emph{Proceedings of the 2006 ACM/IEEE Conference on
  Supercomputing (SC)}}.
\newblock
\href{https://doi.org/10.1145/1188455.1188543}{doi:\nolinkurl{10.1145/1188455.1188543}}


\bibitem[Frigo et~al\mbox{.}(1998)]%
        {frigo1998implementation}
\bibfield{author}{\bibinfo{person}{Matteo Frigo}, \bibinfo{person}{Charles~E.
  Leiserson}, {and} \bibinfo{person}{Keith~H. Randall}.}
  \bibinfo{year}{1998}\natexlab{}.
\newblock \showarticletitle{The Implementation of the {Cilk-5} Multithreaded
  Language}. In \bibinfo{booktitle}{\emph{Proceedings of the ACM SIGPLAN 1998
  Conference on Programming Language Design and Implementation (PLDI)}}.
  \bibinfo{publisher}{ACM}, \bibinfo{pages}{212--223}.
\newblock
\href{https://doi.org/10.1145/277650.277725}{doi:\nolinkurl{10.1145/277650.277725}}


\bibitem[Gruevski et~al\mbox{.}(2026)]%
        {cargosemverchecks}
\bibfield{author}{\bibinfo{person}{Predrag Gruevski} {et~al\mbox{.}}}
  \bibinfo{year}{2026}\natexlab{}.
\newblock \bibinfo{title}{cargo-semver-checks: Scan Your {Rust} Crate for
  Semver Violations}.
\newblock
  \bibinfo{howpublished}{\url{https://github.com/obi1kenobi/cargo-semver-checks}}.
\newblock
\newblock
\shownote{Accessed 2026-10-16}.


\bibitem[Henriksen et~al\mbox{.}(2017)]%
        {henriksen2017futhark}
\bibfield{author}{\bibinfo{person}{Troels Henriksen}, \bibinfo{person}{Niels
  G.~W. Serup}, \bibinfo{person}{Martin Elsman}, \bibinfo{person}{Fritz
  Henglein}, {and} \bibinfo{person}{Cosmin~E. Oancea}.}
  \bibinfo{year}{2017}\natexlab{}.
\newblock \showarticletitle{Futhark: Purely Functional {GPU}-Programming with
  Nested Parallelism and In-Place Array Updates}. In
  \bibinfo{booktitle}{\emph{Proceedings of the 38th ACM SIGPLAN Conference on
  Programming Language Design and Implementation (PLDI '17)}}.
  \bibinfo{publisher}{ACM}, \bibinfo{pages}{556--571}.
\newblock
\href{https://doi.org/10.1145/3062341.3062354}{doi:\nolinkurl{10.1145/3062341.3062354}}


\bibitem[Ikarashi et~al\mbox{.}(2022)]%
        {exo}
\bibfield{author}{\bibinfo{person}{Yuka Ikarashi},
  \bibinfo{person}{Gilbert~Louis Bernstein}, \bibinfo{person}{Alex Reinking},
  \bibinfo{person}{Hasan Genc}, {and} \bibinfo{person}{Jonathan Ragan-Kelley}.}
  \bibinfo{year}{2022}\natexlab{}.
\newblock \showarticletitle{Exocompilation for Productive Programming of
  Hardware Accelerators}. In \bibinfo{booktitle}{\emph{Proceedings of the 43rd
  ACM SIGPLAN International Conference on Programming Language Design and
  Implementation (PLDI)}}.
\newblock
\href{https://doi.org/10.1145/3519939.3523446}{doi:\nolinkurl{10.1145/3519939.3523446}}


\bibitem[Jung et~al\mbox{.}(2018)]%
        {jung2018rustbelt}
\bibfield{author}{\bibinfo{person}{Ralf Jung}, \bibinfo{person}{Jacques-Henri
  Jourdan}, \bibinfo{person}{Robbert Krebbers}, {and} \bibinfo{person}{Derek
  Dreyer}.} \bibinfo{year}{2018}\natexlab{}.
\newblock \showarticletitle{{RustBelt}: Securing the Foundations of the {Rust}
  Programming Language}.
\newblock \bibinfo{journal}{\emph{Proceedings of the ACM on Programming
  Languages}} \bibinfo{volume}{2}, \bibinfo{number}{POPL}
  (\bibinfo{year}{2018}), \bibinfo{pages}{66:1--66:34}.
\newblock
\href{https://doi.org/10.1145/3158154}{doi:\nolinkurl{10.1145/3158154}}


\bibitem[Jung et~al\mbox{.}(2026)]%
        {jung2026miri}
\bibfield{author}{\bibinfo{person}{Ralf Jung}, \bibinfo{person}{Benjamin
  Kimock}, \bibinfo{person}{Christian Poveda},
  \bibinfo{person}{Eduardo~S{\'a}nchez Mu{\~n}oz}, \bibinfo{person}{Oli
  Scherer}, {and} \bibinfo{person}{Qian Wang}.}
  \bibinfo{year}{2026}\natexlab{}.
\newblock \showarticletitle{Miri: Practical Undefined Behavior Detection for
  {Rust}}.
\newblock \bibinfo{journal}{\emph{Proceedings of the ACM on Programming
  Languages}} \bibinfo{volume}{10}, \bibinfo{number}{POPL}
  (\bibinfo{year}{2026}), \bibinfo{pages}{1383--1411}.
\newblock
\href{https://doi.org/10.1145/3776690}{doi:\nolinkurl{10.1145/3776690}}


\bibitem[{Khronos SYCL Working Group}({[n.\,d.]})]%
        {sycl2020}
\bibfield{author}{\bibinfo{person}{{Khronos SYCL Working Group}}.}
  \bibinfo{year}{[n.\,d.]}\natexlab{}.
\newblock \bibinfo{title}{{SYCL} 2020 Specification (revision 12)}.
\newblock
  \bibinfo{howpublished}{\url{https://registry.khronos.org/SYCL/specs/sycl-2020/html/sycl-2020.html}}.
\newblock
\newblock
\shownote{Accessed 28 October 2026}.


\bibitem[Kokologiannakis and Vafeiadis(2021)]%
        {kokologiannakis2021genmc}
\bibfield{author}{\bibinfo{person}{Michalis Kokologiannakis} {and}
  \bibinfo{person}{Viktor Vafeiadis}.} \bibinfo{year}{2021}\natexlab{}.
\newblock \showarticletitle{{GenMC}: A Model Checker for Weak Memory Models}.
  In \bibinfo{booktitle}{\emph{Computer Aided Verification (CAV 2021)}}
  \emph{(\bibinfo{series}{Lecture Notes in Computer Science},
  Vol.~\bibinfo{volume}{12759})}. \bibinfo{publisher}{Springer},
  \bibinfo{pages}{427--440}.
\newblock
\href{https://doi.org/10.1007/978-3-030-81685-8_20}{doi:\nolinkurl{10.1007/978-3-030-81685-8_20}}


\bibitem[K{\"o}pcke et~al\mbox{.}(2024)]%
        {descend}
\bibfield{author}{\bibinfo{person}{Bastian K{\"o}pcke}, \bibinfo{person}{Sergei
  Gorlatch}, {and} \bibinfo{person}{Michel Steuwer}.}
  \bibinfo{year}{2024}\natexlab{}.
\newblock \showarticletitle{Descend: A Safe {GPU} Systems Programming
  Language}.
\newblock \bibinfo{journal}{\emph{Proceedings of the ACM on Programming
  Languages}} \bibinfo{volume}{8}, \bibinfo{number}{PLDI}
  (\bibinfo{year}{2024}).
\newblock
\href{https://doi.org/10.1145/3656411}{doi:\nolinkurl{10.1145/3656411}}


\bibitem[Kumar(2026a)]%
        {kumar2026frontend}
\bibfield{author}{\bibinfo{person}{Aditya Kumar}.}
  \bibinfo{year}{2026}\natexlab{a}.
\newblock \bibinfo{title}{Agree Without Locking: A Deterministic Parallel
  Compiler Frontend}.
\newblock
\newblock
\shownote{Preprint}.


\bibitem[Kumar(2026b)]%
        {kumar2026admission}
\bibfield{author}{\bibinfo{person}{Aditya Kumar}.}
  \bibinfo{year}{2026}\natexlab{b}.
\newblock \bibinfo{title}{Compile-Time {GPU} Memory Checks for {LLM} Serving
  Across Heterogeneous Machines}.
\newblock
\newblock
\shownote{Preprint}.


\bibitem[Kumar(2026c)]%
        {kumar2026comptime}
\bibfield{author}{\bibinfo{person}{Aditya Kumar}.}
  \bibinfo{year}{2026}\natexlab{c}.
\newblock \bibinfo{title}{Comptime in {Vx}: Design, Soundness and Testing}.
\newblock
\newblock
\shownote{Technical report}.


\bibitem[Kumar(2026d)]%
        {kumar2026erasable}
\bibfield{author}{\bibinfo{person}{Aditya Kumar}.}
  \bibinfo{year}{2026}\natexlab{d}.
\newblock \bibinfo{title}{Erasable Parallelism: {GPU} Loop Nests That Keep
  Their Sequential Meaning}.
\newblock
\newblock
\shownote{Preprint}.


\bibitem[Kumar(2026e)]%
        {kumar2026machinemodel}
\bibfield{author}{\bibinfo{person}{Aditya Kumar}.}
  \bibinfo{year}{2026}\natexlab{e}.
\newblock \bibinfo{title}{A Machine Model That Carries Provenance in a
  Heterogeneous Systems Programming Language}.
\newblock
\newblock
\shownote{Preprint}.


\bibitem[Kumar(2026f)]%
        {kumar2026mustfold}
\bibfield{author}{\bibinfo{person}{Aditya Kumar}.}
  \bibinfo{year}{2026}\natexlab{f}.
\newblock \bibinfo{title}{Must-Fold: Compile-Time Blocks That Can See Run-Time
  Variables}.
\newblock
\newblock
\shownote{Preprint}.


\bibitem[Kumar(2026g)]%
        {kumar2026tasks}
\bibfield{author}{\bibinfo{person}{Aditya Kumar}.}
  \bibinfo{year}{2026}\natexlab{g}.
\newblock \bibinfo{title}{Structured Tasks and Footprints: Data-Race Freedom
  for a Systems Language, Implemented and Mechanized}.
\newblock
\newblock
\shownote{Preprint}.


\bibitem[Kumar(2026h)]%
        {kumar2026fleet}
\bibfield{author}{\bibinfo{person}{Aditya Kumar}.}
  \bibinfo{year}{2026}\natexlab{h}.
\newblock \bibinfo{title}{{Vx}: A Programming Language for Fleet-Level
  Programming}.
\newblock
\newblock
\shownote{Technical report}.


\bibitem[Kumar(2026i)]%
        {vx2026}
\bibfield{author}{\bibinfo{person}{Aditya Kumar}.}
  \bibinfo{year}{2026}\natexlab{i}.
\newblock \bibinfo{title}{{Vx}: a systems programming language for
  heterogeneous computing}.
\newblock \bibinfo{howpublished}{\url{https://vxlang.org}}.
\newblock
\newblock
\shownote{Version 0.0.5. Source at \url{https://github.com/vx-lang/Vx}}.


\bibitem[Lahav et~al\mbox{.}(2017)]%
        {lahav2017rc11}
\bibfield{author}{\bibinfo{person}{Ori Lahav}, \bibinfo{person}{Viktor
  Vafeiadis}, \bibinfo{person}{Jeehoon Kang}, \bibinfo{person}{Chung-Kil Hur},
  {and} \bibinfo{person}{Derek Dreyer}.} \bibinfo{year}{2017}\natexlab{}.
\newblock \showarticletitle{Repairing Sequential Consistency in {C/C++11}}. In
  \bibinfo{booktitle}{\emph{Proceedings of the 38th ACM SIGPLAN Conference on
  Programming Language Design and Implementation (PLDI)}}.
  \bibinfo{pages}{618--632}.
\newblock
\href{https://doi.org/10.1145/3062341.3062352}{doi:\nolinkurl{10.1145/3062341.3062352}}


\bibitem[Lattner and Adve(2004)]%
        {llvm}
\bibfield{author}{\bibinfo{person}{Chris Lattner} {and} \bibinfo{person}{Vikram
  Adve}.} \bibinfo{year}{2004}\natexlab{}.
\newblock \showarticletitle{{LLVM}: A Compilation Framework for Lifelong
  Program Analysis and Transformation}. In
  \bibinfo{booktitle}{\emph{International Symposium on Code Generation and
  Optimization (CGO)}}. \bibinfo{pages}{75--86}.
\newblock
\href{https://doi.org/10.1109/CGO.2004.1281665}{doi:\nolinkurl{10.1109/CGO.2004.1281665}}


\bibitem[Lattner et~al\mbox{.}(2021)]%
        {mlir}
\bibfield{author}{\bibinfo{person}{Chris Lattner}, \bibinfo{person}{Mehdi
  Amini}, \bibinfo{person}{Uday Bondhugula}, \bibinfo{person}{Albert Cohen},
  \bibinfo{person}{Andy Davis}, \bibinfo{person}{Jacques Pienaar},
  \bibinfo{person}{River Riddle}, \bibinfo{person}{Tatiana Shpeisman},
  \bibinfo{person}{Nicolas Vasilache}, {and} \bibinfo{person}{Oleksandr
  Zinenko}.} \bibinfo{year}{2021}\natexlab{}.
\newblock \showarticletitle{{MLIR}: Scaling Compiler Infrastructure for Domain
  Specific Computation}. In \bibinfo{booktitle}{\emph{Proceedings of the 2021
  IEEE/ACM International Symposium on Code Generation and Optimization (CGO)}}.
  \bibinfo{publisher}{IEEE}, \bibinfo{pages}{2--14}.
\newblock
\href{https://doi.org/10.1109/CGO51591.2021.9370308}{doi:\nolinkurl{10.1109/CGO51591.2021.9370308}}


\bibitem[Lattuada et~al\mbox{.}(2023)]%
        {lattuada2023verus}
\bibfield{author}{\bibinfo{person}{Andrea Lattuada}, \bibinfo{person}{Travis
  Hance}, \bibinfo{person}{Chanhee Cho}, \bibinfo{person}{Matthias Brun},
  \bibinfo{person}{Isitha Subasinghe}, \bibinfo{person}{Yi Zhou},
  \bibinfo{person}{Jon Howell}, \bibinfo{person}{Bryan Parno}, {and}
  \bibinfo{person}{Chris Hawblitzel}.} \bibinfo{year}{2023}\natexlab{}.
\newblock \showarticletitle{{Verus}: Verifying {Rust} Programs using Linear
  Ghost Types}.
\newblock \bibinfo{journal}{\emph{Proceedings of the ACM on Programming
  Languages}} \bibinfo{volume}{7}, \bibinfo{number}{OOPSLA1}
  (\bibinfo{year}{2023}), \bibinfo{pages}{286--315}.
\newblock
\href{https://doi.org/10.1145/3586037}{doi:\nolinkurl{10.1145/3586037}}


\bibitem[Leino(2010)]%
        {leino2010dafny}
\bibfield{author}{\bibinfo{person}{K.~Rustan~M. Leino}.}
  \bibinfo{year}{2010}\natexlab{}.
\newblock \showarticletitle{Dafny: An Automatic Program Verifier for Functional
  Correctness}. In \bibinfo{booktitle}{\emph{Logic for Programming, Artificial
  Intelligence, and Reasoning (LPAR)}}. \bibinfo{publisher}{Springer},
  \bibinfo{pages}{348--370}.
\newblock
\href{https://doi.org/10.1007/978-3-642-17511-4_20}{doi:\nolinkurl{10.1007/978-3-642-17511-4_20}}


\bibitem[Leroy(2009)]%
        {leroy2009compcert}
\bibfield{author}{\bibinfo{person}{Xavier Leroy}.}
  \bibinfo{year}{2009}\natexlab{}.
\newblock \showarticletitle{Formal Verification of a Realistic Compiler}.
\newblock \bibinfo{journal}{\emph{Commun. ACM}} \bibinfo{volume}{52},
  \bibinfo{number}{7} (\bibinfo{year}{2009}), \bibinfo{pages}{107--115}.
\newblock
\href{https://doi.org/10.1145/1538788.1538814}{doi:\nolinkurl{10.1145/1538788.1538814}}


\bibitem[Matsakis and Klock(2014)]%
        {matsakis2014rust}
\bibfield{author}{\bibinfo{person}{Nicholas~D. Matsakis} {and}
  \bibinfo{person}{Felix~S. Klock}.} \bibinfo{year}{2014}\natexlab{}.
\newblock \showarticletitle{The {Rust} Language}.
\newblock \bibinfo{journal}{\emph{ACM SIGAda Ada Letters}}
  \bibinfo{volume}{34}, \bibinfo{number}{3} (\bibinfo{year}{2014}),
  \bibinfo{pages}{103--104}.
\newblock
\href{https://doi.org/10.1145/2692956.2663188}{doi:\nolinkurl{10.1145/2692956.2663188}}


\bibitem[McKeeman(1998)]%
        {mckeeman1998differential}
\bibfield{author}{\bibinfo{person}{William~M. McKeeman}.}
  \bibinfo{year}{1998}\natexlab{}.
\newblock \showarticletitle{Differential Testing for Software}.
\newblock \bibinfo{journal}{\emph{Digital Technical Journal}}
  \bibinfo{volume}{10}, \bibinfo{number}{1} (\bibinfo{year}{1998}),
  \bibinfo{pages}{100--107}.
\newblock


\bibitem[Melicher et~al\mbox{.}(2017)]%
        {melicher2017capability}
\bibfield{author}{\bibinfo{person}{Darya Melicher},
  \bibinfo{person}{Yangqingwei Shi}, \bibinfo{person}{Alex Potanin}, {and}
  \bibinfo{person}{Jonathan Aldrich}.} \bibinfo{year}{2017}\natexlab{}.
\newblock \showarticletitle{A Capability-Based Module System for Authority
  Control}. In \bibinfo{booktitle}{\emph{31st European Conference on
  Object-Oriented Programming (ECOOP)}} \emph{(\bibinfo{series}{LIPIcs},
  Vol.~\bibinfo{volume}{74})}. \bibinfo{pages}{20:1--20:27}.
\newblock
\href{https://doi.org/10.4230/LIPIcs.ECOOP.2017.20}{doi:\nolinkurl{10.4230/LIPIcs.ECOOP.2017.20}}


\bibitem[Mettler et~al\mbox{.}(2010)]%
        {mettler2010joee}
\bibfield{author}{\bibinfo{person}{Adrian Mettler}, \bibinfo{person}{David
  Wagner}, {and} \bibinfo{person}{Tyler Close}.}
  \bibinfo{year}{2010}\natexlab{}.
\newblock \showarticletitle{{Joe-E}: A Security-Oriented Subset of {Java}}. In
  \bibinfo{booktitle}{\emph{Network and Distributed System Security Symposium
  (NDSS)}}.
\newblock
\urldef\tempurl%
\url{https://www.ndss-symposium.org/ndss2010/joe-e-security-oriented-subset-java/}
\showURL{%
\tempurl}


\bibitem[Miller(2006)]%
        {miller2006robust}
\bibfield{author}{\bibinfo{person}{Mark~Samuel Miller}.}
  \bibinfo{year}{2006}\natexlab{}.
\newblock \emph{\bibinfo{title}{Robust Composition: Towards a Unified Approach
  to Access Control and Concurrency Control}}.
\newblock \bibinfo{thesistype}{Ph.\,D. Dissertation}. \bibinfo{school}{Johns
  Hopkins University}.
\newblock


\bibitem[{Modular Inc.}(2023)]%
        {mojo}
\bibfield{author}{\bibinfo{person}{{Modular Inc.}}}
  \bibinfo{year}{2023}\natexlab{}.
\newblock \bibinfo{title}{The Mojo Programming Language}.
\newblock \bibinfo{howpublished}{\url{https://www.modular.com/mojo}}.
\newblock


\bibitem[Moses and Churavy(2020)]%
        {moses2020enzyme}
\bibfield{author}{\bibinfo{person}{William~S. Moses} {and}
  \bibinfo{person}{Valentin Churavy}.} \bibinfo{year}{2020}\natexlab{}.
\newblock \showarticletitle{Instead of Rewriting Foreign Code for Machine
  Learning, Automatically Synthesize Fast Gradients}. In
  \bibinfo{booktitle}{\emph{Advances in Neural Information Processing Systems
  33 (NeurIPS)}}.
\newblock
\newblock
\shownote{arXiv:2010.01709}.


\bibitem[{NVIDIA Corporation}(2026)]%
        {cudaguide}
\bibfield{author}{\bibinfo{person}{{NVIDIA Corporation}}.}
  \bibinfo{year}{2026}\natexlab{}.
\newblock \bibinfo{title}{{CUDA} {C++} Programming Guide}.
\newblock
  \bibinfo{howpublished}{\url{https://docs.nvidia.com/cuda/cuda-c-programming-guide/}}.
\newblock
\newblock
\shownote{Accessed 23 October 2026}.


\bibitem[Ragan-Kelley et~al\mbox{.}(2013)]%
        {ragankelley13}
\bibfield{author}{\bibinfo{person}{Jonathan Ragan-Kelley},
  \bibinfo{person}{Connelly Barnes}, \bibinfo{person}{Andrew Adams},
  \bibinfo{person}{Sylvain Paris}, \bibinfo{person}{Fr{\'e}do Durand}, {and}
  \bibinfo{person}{Saman Amarasinghe}.} \bibinfo{year}{2013}\natexlab{}.
\newblock \showarticletitle{Halide: A Language and Compiler for Optimizing
  Parallelism, Locality, and Recomputation in Image Processing Pipelines}. In
  \bibinfo{booktitle}{\emph{PLDI}}.
\newblock


\bibitem[Shali and Cook(2011)]%
        {shali2011hybrid}
\bibfield{author}{\bibinfo{person}{Amin Shali} {and}
  \bibinfo{person}{William~R. Cook}.} \bibinfo{year}{2011}\natexlab{}.
\newblock \showarticletitle{Hybrid Partial Evaluation}. In
  \bibinfo{booktitle}{\emph{Proceedings of the 2011 ACM International
  Conference on Object Oriented Programming Systems Languages and Applications
  (OOPSLA)}}. \bibinfo{publisher}{ACM}, \bibinfo{address}{New York, NY, USA},
  \bibinfo{pages}{375--390}.
\newblock
\href{https://doi.org/10.1145/2048066.2048098}{doi:\nolinkurl{10.1145/2048066.2048098}}


\bibitem[Slaughter et~al\mbox{.}(2015)]%
        {regent}
\bibfield{author}{\bibinfo{person}{Elliott Slaughter}, \bibinfo{person}{Wonchan
  Lee}, \bibinfo{person}{Sean Treichler}, \bibinfo{person}{Michael Bauer},
  {and} \bibinfo{person}{Alex Aiken}.} \bibinfo{year}{2015}\natexlab{}.
\newblock \showarticletitle{Regent: A High-Productivity Programming Language
  for {HPC} with Logical Regions}. In \bibinfo{booktitle}{\emph{Proceedings of
  the International Conference for High Performance Computing, Networking,
  Storage and Analysis (SC)}}.
\newblock
\href{https://doi.org/10.1145/2807591.2807629}{doi:\nolinkurl{10.1145/2807591.2807629}}


\bibitem[Stone et~al\mbox{.}(2010)]%
        {opencl}
\bibfield{author}{\bibinfo{person}{John~E. Stone}, \bibinfo{person}{David
  Gohara}, {and} \bibinfo{person}{Guochun Shi}.}
  \bibinfo{year}{2010}\natexlab{}.
\newblock \showarticletitle{{OpenCL}: A Parallel Programming Standard for
  Heterogeneous Computing Systems}.
\newblock \bibinfo{journal}{\emph{Computing in Science \& Engineering}}
  \bibinfo{volume}{12}, \bibinfo{number}{3} (\bibinfo{year}{2010}).
\newblock
\href{https://doi.org/10.1109/MCSE.2010.69}{doi:\nolinkurl{10.1109/MCSE.2010.69}}


\bibitem[Swamy et~al\mbox{.}(2016)]%
        {swamy2016fstar}
\bibfield{author}{\bibinfo{person}{Nikhil Swamy},
  \bibinfo{person}{C{\u{a}}t{\u{a}}lin Hri{\c{t}}cu}, \bibinfo{person}{Chantal
  Keller}, \bibinfo{person}{Aseem Rastogi}, \bibinfo{person}{Antoine
  Delignat-Lavaud}, \bibinfo{person}{Simon Forest},
  \bibinfo{person}{Karthikeyan Bhargavan}, \bibinfo{person}{C{\'e}dric
  Fournet}, \bibinfo{person}{Pierre-Yves Strub}, \bibinfo{person}{Markulf
  Kohlweiss}, \bibinfo{person}{Jean-Karim Zinzindohoue}, {and}
  \bibinfo{person}{Santiago Zanella-B{\'e}guelin}.}
  \bibinfo{year}{2016}\natexlab{}.
\newblock \showarticletitle{Dependent Types and Multi-Monadic Effects in {F*}}.
  In \bibinfo{booktitle}{\emph{Proceedings of the 43rd Annual ACM
  SIGPLAN-SIGACT Symposium on Principles of Programming Languages (POPL)}}.
  \bibinfo{publisher}{ACM}, \bibinfo{pages}{256--270}.
\newblock
\href{https://doi.org/10.1145/2837614.2837655}{doi:\nolinkurl{10.1145/2837614.2837655}}


\bibitem[Taha and Sheard(1997)]%
        {taha1997multistage}
\bibfield{author}{\bibinfo{person}{Walid Taha} {and} \bibinfo{person}{Tim
  Sheard}.} \bibinfo{year}{1997}\natexlab{}.
\newblock \showarticletitle{Multi-Stage Programming with Explicit Annotations}.
  In \bibinfo{booktitle}{\emph{Proceedings of the 1997 ACM SIGPLAN Symposium on
  Partial Evaluation and Semantics-Based Program Manipulation (PEPM)}}.
  \bibinfo{publisher}{ACM}, \bibinfo{address}{New York, NY, USA},
  \bibinfo{pages}{203--217}.
\newblock
\href{https://doi.org/10.1145/258993.259019}{doi:\nolinkurl{10.1145/258993.259019}}


\bibitem[{The Carbon Language project}(2026)]%
        {carbon}
\bibfield{author}{\bibinfo{person}{{The Carbon Language project}}.}
  \bibinfo{year}{2026}\natexlab{}.
\newblock \bibinfo{title}{Carbon Language: An experimental successor to {C++}}.
\newblock \bibinfo{howpublished}{GitHub repository}.
\newblock
\urldef\tempurl%
\url{https://github.com/carbon-language/carbon-lang}
\showURL{%
\tempurl}
\newblock
\shownote{Accessed August 2026}.


\bibitem[{The Pony Developers}(2026)]%
        {ponycaps}
\bibfield{author}{\bibinfo{person}{{The Pony Developers}}.}
  \bibinfo{year}{2026}\natexlab{}.
\newblock \bibinfo{title}{Object Capabilities}.
\newblock \bibinfo{howpublished}{Pony tutorial,
  \url{https://tutorial.ponylang.io/object-capabilities/object-capabilities.html}}.
\newblock
\newblock
\shownote{Accessed 2026-11-24}.


\bibitem[Tillet et~al\mbox{.}(2019)]%
        {tillet19}
\bibfield{author}{\bibinfo{person}{Philippe Tillet}, \bibinfo{person}{H.~T.
  Kung}, {and} \bibinfo{person}{David Cox}.} \bibinfo{year}{2019}\natexlab{}.
\newblock \showarticletitle{Triton: An Intermediate Language and Compiler for
  Tiled Neural Network Computations}. In \bibinfo{booktitle}{\emph{MAPL}}.
\newblock


\bibitem[VanHattum et~al\mbox{.}(2022)]%
        {vanhattum2022kani}
\bibfield{author}{\bibinfo{person}{Alexa VanHattum}, \bibinfo{person}{Daniel
  Schwartz-Narbonne}, \bibinfo{person}{Nathan Chong}, {and}
  \bibinfo{person}{Adrian Sampson}.} \bibinfo{year}{2022}\natexlab{}.
\newblock \showarticletitle{Verifying Dynamic Trait Objects in {Rust}}. In
  \bibinfo{booktitle}{\emph{Proceedings of the 44th International Conference on
  Software Engineering: Software Engineering in Practice (ICSE-SEIP)}}.
  \bibinfo{publisher}{ACM}, \bibinfo{pages}{321--330}.
\newblock
\href{https://doi.org/10.1145/3510457.3513031}{doi:\nolinkurl{10.1145/3510457.3513031}}


\bibitem[Vazou et~al\mbox{.}(2014)]%
        {vazou2014refinement}
\bibfield{author}{\bibinfo{person}{Niki Vazou}, \bibinfo{person}{Eric~L.
  Seidel}, \bibinfo{person}{Ranjit Jhala}, \bibinfo{person}{Dimitrios
  Vytiniotis}, {and} \bibinfo{person}{Simon~Peyton Jones}.}
  \bibinfo{year}{2014}\natexlab{}.
\newblock \showarticletitle{Refinement Types for {Haskell}}. In
  \bibinfo{booktitle}{\emph{Proceedings of the 19th ACM SIGPLAN International
  Conference on Functional Programming (ICFP)}}. \bibinfo{publisher}{ACM},
  \bibinfo{pages}{269--282}.
\newblock
\href{https://doi.org/10.1145/2628136.2628161}{doi:\nolinkurl{10.1145/2628136.2628161}}


\bibitem[VII et~al\mbox{.}(2008)]%
        {ml5}
\bibfield{author}{\bibinfo{person}{Tom~Murphy VII}, \bibinfo{person}{Karl
  Crary}, {and} \bibinfo{person}{Robert Harper}.}
  \bibinfo{year}{2008}\natexlab{}.
\newblock \showarticletitle{Type-Safe Distributed Programming with {ML5}}. In
  \bibinfo{booktitle}{\emph{Trustworthy Global Computing (TGC)}}.
\newblock
\href{https://doi.org/10.1007/978-3-540-78663-4_9}{doi:\nolinkurl{10.1007/978-3-540-78663-4_9}}


\bibitem[VII et~al\mbox{.}(2004)]%
        {lambda5}
\bibfield{author}{\bibinfo{person}{Tom~Murphy VII}, \bibinfo{person}{Karl
  Crary}, \bibinfo{person}{Robert Harper}, {and} \bibinfo{person}{Frank
  Pfenning}.} \bibinfo{year}{2004}\natexlab{}.
\newblock \showarticletitle{A Symmetric Modal Lambda Calculus for Distributed
  Computing}. In \bibinfo{booktitle}{\emph{Proceedings of the 19th IEEE
  Symposium on Logic in Computer Science (LICS)}}.
\newblock


\bibitem[Watson et~al\mbox{.}(2010)]%
        {watson2010capsicum}
\bibfield{author}{\bibinfo{person}{Robert N.~M. Watson},
  \bibinfo{person}{Jonathan Anderson}, \bibinfo{person}{Ben Laurie}, {and}
  \bibinfo{person}{Kris Kennaway}.} \bibinfo{year}{2010}\natexlab{}.
\newblock \showarticletitle{Capsicum: Practical Capabilities for {UNIX}}. In
  \bibinfo{booktitle}{\emph{19th USENIX Security Symposium}}.
\newblock
\urldef\tempurl%
\url{https://www.usenix.org/conference/usenixsecurity10/capsicum-practical-capabilities-unix}
\showURL{%
\tempurl}


\bibitem[{WebAssembly Community Group}(2026)]%
        {wasi}
\bibfield{author}{\bibinfo{person}{{WebAssembly Community Group}}.}
  \bibinfo{year}{2026}\natexlab{}.
\newblock \bibinfo{title}{{WASI}: The {WebAssembly} System Interface}.
\newblock \bibinfo{howpublished}{\url{https://wasi.dev/}}.
\newblock
\newblock
\shownote{Accessed 2026-11-24}.


\bibitem[Wickerson et~al\mbox{.}(2017)]%
        {wickerson2017memalloy}
\bibfield{author}{\bibinfo{person}{John Wickerson}, \bibinfo{person}{Mark
  Batty}, \bibinfo{person}{Tyler Sorensen}, {and} \bibinfo{person}{George~A.
  Constantinides}.} \bibinfo{year}{2017}\natexlab{}.
\newblock \showarticletitle{Automatically Comparing Memory Consistency Models}.
  In \bibinfo{booktitle}{\emph{Proceedings of the 44th ACM SIGPLAN Symposium on
  Principles of Programming Languages (POPL)}}. \bibinfo{pages}{190--204}.
\newblock
\href{https://doi.org/10.1145/3009837.3009838}{doi:\nolinkurl{10.1145/3009837.3009838}}


\bibitem[Yang et~al\mbox{.}(2011)]%
        {yang2011csmith}
\bibfield{author}{\bibinfo{person}{Xuejun Yang}, \bibinfo{person}{Yang Chen},
  \bibinfo{person}{Eric Eide}, {and} \bibinfo{person}{John Regehr}.}
  \bibinfo{year}{2011}\natexlab{}.
\newblock \showarticletitle{Finding and Understanding Bugs in {C} Compilers}.
  In \bibinfo{booktitle}{\emph{Proceedings of the 32nd ACM SIGPLAN Conference
  on Programming Language Design and Implementation (PLDI)}}.
  \bibinfo{publisher}{ACM}, \bibinfo{address}{New York, NY, USA},
  \bibinfo{pages}{283--294}.
\newblock
\href{https://doi.org/10.1145/1993498.1993532}{doi:\nolinkurl{10.1145/1993498.1993532}}


\bibitem[{Zig Software Foundation}(2026)]%
        {zigdocs}
\bibfield{author}{\bibinfo{person}{{Zig Software Foundation}}.}
  \bibinfo{year}{2026}\natexlab{}.
\newblock \bibinfo{title}{Zig Language Reference: Compile-Time Expressions}.
\newblock
  \bibinfo{howpublished}{\url{https://ziglang.org/documentation/master/}}.
\newblock
\newblock
\shownote{Behaviour checked with Zig 0.17.0}.


\end{thebibliography}

\appendix
\section{Vx in One Page}\label{app:quickref}

This appendix summarizes the syntax of Vx at \kw{719726333}, after the quick reference of the Vx
book. Each line is a form that the compiler accepts.

\begin{table}[h]
\small
\begin{tabularx}{\linewidth}{@{}p{0.2\linewidth}Y@{}}
\toprule
Part & Forms \\
\midrule
Program & One file is one module; \kw{import core::option;} brings in what a module declares;
  \kw{fn main() -> i32} starts the program, and its result is the exit code. \\
Declarations & \kw{const N : i64 = 4;} \quad \kw{fn add(a : i64, b : i64) -> i64 \{ .. \}} \quad
  \kw{struct Point \{ x : i64, y : i64 \}} \quad \kw{enum Shape \{ Square(i64), Rect(i64, i64) \}} \quad
  \kw{trait Describe \{ fn size(self : \&Self) -> i64; \}} \quad \kw{impl Describe for Shape \{ .. \}} in the file of \kw{Shape}; \kw{extern impl} in another file \\
Generics & \kw{struct Pair<T>} \quad \kw{fn largest<T : Describe>(a : \&T) -> i64} \quad
  \kw{fn dot<const N : i32>(a : Tensor<f32, [N]>) -> f32}; a generic type always writes its arguments:
  \kw{Vec<i64>::new()}, \kw{Option<i64>::Some(3)}. \\
Numbers & \kw{i4}--\kw{i64}, \kw{u4}--\kw{u64}, \kw{usize}, \kw{f16}, \kw{bf16}, \kw{f32},
  \kw{f64}, \kw{bool}, eight- and four-bit floats; no implicit conversion (\kw{x as i64}); \kw{+ - *}
  wrap; \kw{/} by zero and \kw{usize} overflow stop the program. \\
Data & tuples \kw{(i64, f32)} and \kw{let (a, b) = t;}; references \kw{\&T}, \kw{\&mut T}; raw
  pointers \kw{*const T}, \kw{*mut T in Memory::X}; tensors \kw{Tensor<f32, [2, 3]>},
  \kw{Tensor<f32, [?, ?]>}, \kw{Tensor<f32, [4], Memory::HBM>}; \kw{Option}, \kw{Result}, \kw{Vec},
  \kw{Box}. \\
Statements & \kw{let x = 1;} \quad \kw{let mut y : i64 = 2;} \quad \kw{if c \{ .. \} else \{ .. \}}
  (also an expression) \quad \kw{for i in 0..n} \quad \kw{while} \quad \kw{loop} \quad
  \kw{match v \{ Shape::Square(s) => \{ .. \}, \_ => \{ .. \} \}} \quad \kw{let v = parse(s)?;} \\
Closures & \kw{|x : i64| x * 2}; parameter types are written; the capture mode follows the body. \\
Machines & \kw{Memory HBM \{ capacity: 80 GiB, bandwidth: 3.35 TB/s, \ldots \}} \quad
  \kw{Topology Device \{ memory: \ldots, visible: [\ldots], levels: [\ldots], transfer A -> B : 63 GB/s \}} \\
Placement & \kw{transfer(x, Memory::HBM)} \quad \kw{transfer\_async(x, M).wait()} \quad
  \kw{spawn on(Topology::Device) \{ .. \}} \\
GPU loops & \kw{for y in 0..64 across blocks \{ for x in 0..300 across lanes(256) \{ .. \} \}};
  \kw{lanes::sum(\&part, n)} between lane loops. \\
Tasks & \kw{let r = tasks \{ let h = spawn \{ .. \}; h.join() \};} \quad
  \kw{let (tx, rx) = channel<i64>(n);} \quad \kw{select2(\&jobs, \&quiet)} gives the index of the first
  ready source \quad \kw{Mutex}, \kw{Semaphore}, \kw{Arc}, atomics. \\
Compile time & \kw{comptime \{ fact(10) \}} \quad \kw{if comptime \ldots} \quad
  \kw{macro\_rules name \{ (\$x : expr) => \{ .. \}; \}}, called as \kw{name!(..)}. \\
Contracts & \kw{requires x < 100} \quad \kw{ensures return > x} \quad loop \kw{invariant}s \quad
  \kw{struct Interval \{ lo : i64, hi : i64, invariant lo <= hi, \}} \quad \kw{assert(c)} \quad borrow clauses \kw{ensures !reaches(return, *k)}
  \quad pointer clauses \kw{requires valid(p, n)}. \\
Unsafe and FFI & \kw{unsafe \{ .. \}} \quad \kw{unsafe fn} \quad \kw{unsafe impl Send for T} (in another file, \kw{unsafe extern impl})
  \quad \kw{assume valid(p, n);} \quad \kw{extern "C" \{ fn abs(x : i32) -> i32; \}} \quad
  \kw{safe fn} in an \kw{extern} block \quad \kw{pub fn} for export \quad \kw{mlir!(\ldots)}. \\
Output & \kw{print(x)} \quad \kw{println!("\{\} and \{\}", a, b)} \quad \kw{eprintln!}
  \quad \kw{panic("..")} \quad \kw{abort()}. \\
Compiler & \kw{vxc file.vx} (compile and run) \quad \kw{--machine F --host default}
  \quad \kw{--action emit-mlir | emit-llvm | emit-obj} \quad \kw{-j N} \quad
  \kw{--diagnostics-json} \quad \kw{--audit} \quad \kw{--verify-seams} \quad
  \kw{--print-borrow-clauses} \quad \kw{-fbounds-checks} \quad \kw{--sanitize=thread}. \\
\bottomrule
\end{tabularx}
\end{table}

\section{How the Facts in This Report Were Found}\label{app:method}

Three readers went through the repository at \kw{a06269cad}, a commit of 20 November 2026. A
second pass checked each fact against main at \kw{719726333}. One read the core language, one the
heterogeneous and parallel features, and one the tools and the project. Each one counted from tracked files and read the code and the tests behind each claim of the
documentation. Each one also listed the places where the documentation claims more than the code. We then built the compiler from a clone of \kw{719726333}: a release build with Rust 1.98 and LLVM 22,
on an Apple M-series computer. We ran every listing, every quoted message and every probe of
Table~\ref{tab:gaps} on it. The admission verdicts of Table~\ref{tab:admit}
come from the same build. The programs are kept with the source of this report. One result in Section~\ref{sec:scalable}
comes from an older 0.0.4 build, and the text says so: a deadlock that the watcher did not
report. Hardware results that we did not reproduce, such as the GPU runs, come from the companion papers or
from the records in the repository. The text says so at each one.

\end{document}